\documentclass[onecolumn,showpacs,showkeys,prb,preprint]{revtex4}
\usepackage[sort&compress]{natbib}
\usepackage{graphicx}
\usepackage{dcolumn}
\usepackage{bm}
\usepackage{subfigure}       
\usepackage{hyperref}

\begin{document}
\title{Three-dimensional Ginzburg-Landau simulation of a vortex line displaced by a zigzag of pinning spheres}

\author{Mauro M. Doria}
 \email{mmd@if.ufrj.br}
\author{Antonio R. de C. Romaguera}%
 \email{ton@if.ufrj.br}
\affiliation{%
Instituto de F\'{\i}sica, Universidade
Federal do Rio de Janeiro\\
C.P. 68528, 21941-972, Rio de Janeiro RJ, Brazil
}%

\author{Welles A. M. Morgado}
\email{welles@if.puc-rj.br}
\affiliation{%
Instituto de F\'{\i}sica, Pontif\'{\i}cia Universidade
Cat\'olica do Rio de Janeiro\\
C.P. 38071, 22452-970, Rio de Janeiro RJ, Brazil
}%

\begin{abstract}{ A vortex line is shaped by a zigzag of pinning
centers and we study here how far the stretched vortex line is able
to follow this path. The pinning center is described by an
insulating sphere of coherence length size such that in its surface
the de Gennes boundary condition applies. We calculate the free
energy density of this system in the framework of the
Ginzburg-Landau theory and study the critical displacement beyond
which the vortex line is detached from the pinning center.}
\end{abstract}

\keywords{{Ginzburg-Landau},{Tridimensional},{pinning theory}}

\pacs{{74.80.-g} {Spatially inhomogeneous structures},{74.25.-q}
{General properties; correlations between physical properties in
normal and superconducting states},{74.20.De} {Phenomenological
theories (two-fluid, Ginzburg-Landau, etc.)}  }

\maketitle
\section{Introduction}
\label{Introduction}

The basic idea of pinning theory is that the vortex is not rigid but
adjustable to a local distribution of
defects\cite{L70,B95,KetSongbook}. In this paper we consider a
vortex line pinned by a zigzag of pinning centers that are
insulating spheres of radius $R$ of the order of the coherence
length $\xi$. The de Gennes boundary condition \cite{deGennesbook}
applies at the pinning sphere surface, considered as an
insulator-superconductor interface. The straight line connecting
these pinning centers is a path with abrupt right and left turns,
whose zigzag fits into a single plane. As pointed out by Yu. N.
Ovchinnikov\cite{OVCH} long ago, the investigation of various types
of inclusions in superconducting materials is of particular interest
because of the onset of metastable states. We have already
considered a similar vortex depinning transition\cite{RD04} to the
one considered here which is being reviewed from the point of view
of displaced defects.
\begin{figure}[!h]
\centering
\includegraphics[width=0.5\linewidth]{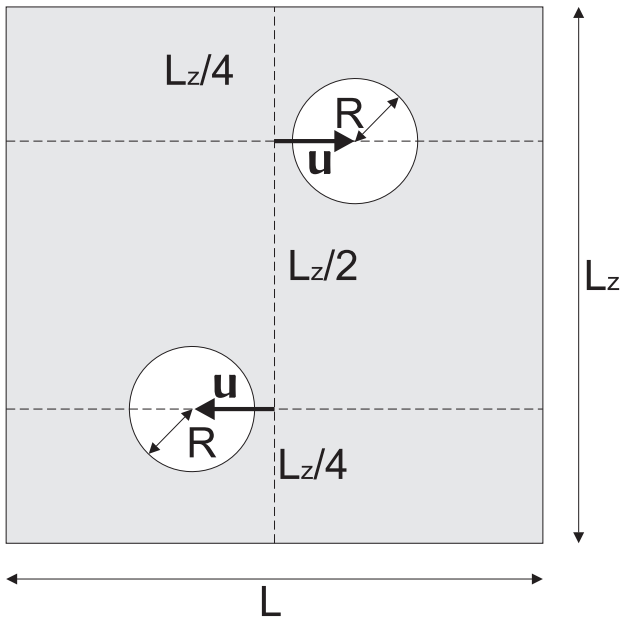}
\caption{Schematic view of the unit cell middle plane. This view
captures the displaced pinning spheres of radius $R$ displaced by
$u$ along the $x$ direction. The unit cell is orthorhombic with base
side $L$ and height $L_z$.} \label{ucellsketch}
\end{figure}

An energetic balance makes the vortex line follows this zigzag path
as long as trapping by the pinning spheres is advantageous as
compared to the increase in length caused by the zigzag path. There
is a critical path that sets a depinning transition beyond which the
stretched vortex line is no longer able to follow the zigzag path of
the pinning centers.

\begin{figure}[!t]
\centering
\includegraphics[width=0.5\linewidth]{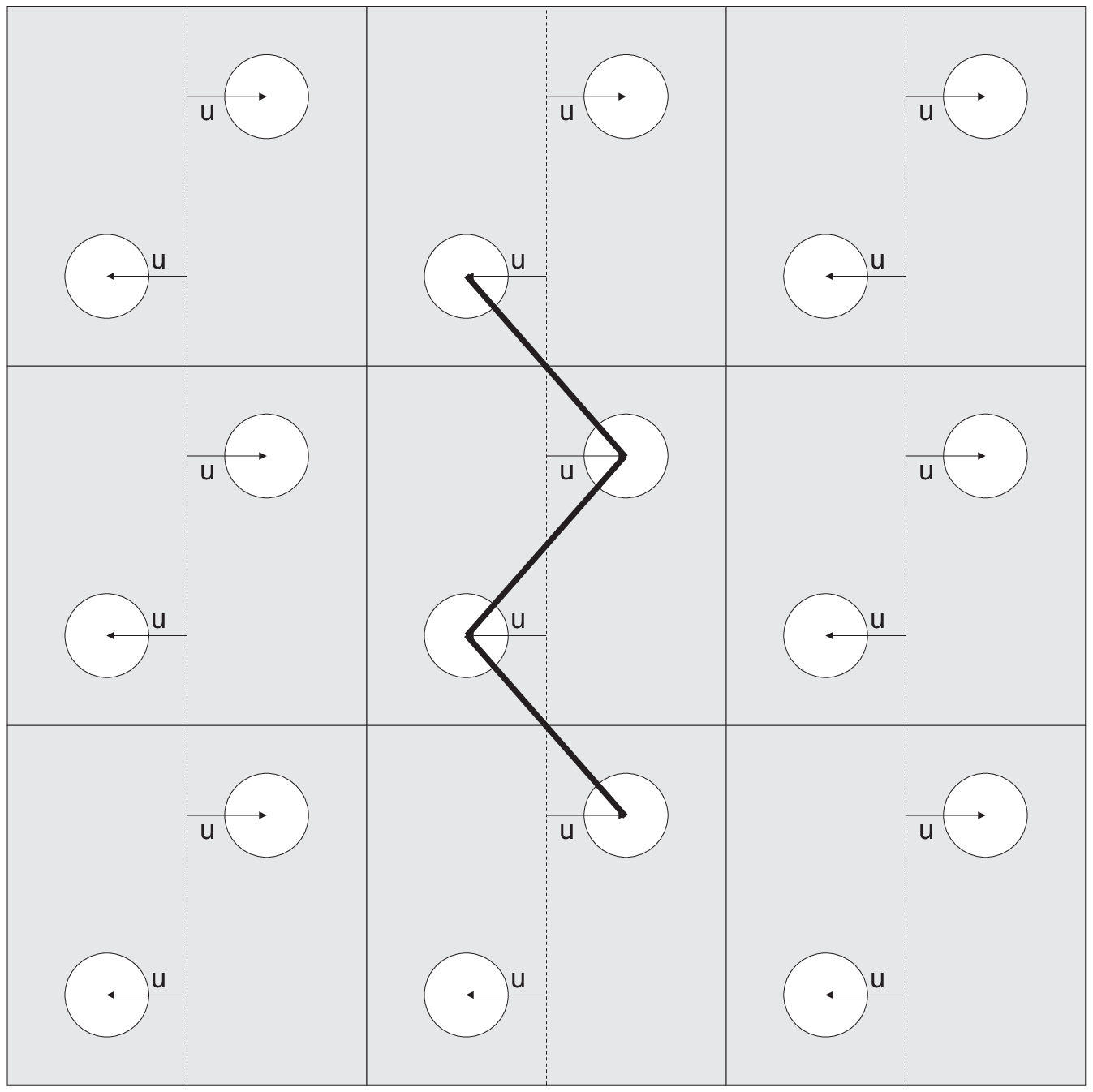}
\caption{Nine equivalent unit cells are patched together here and
the dark line connecting some pinning spheres shows the zigzag path
followed by the vortex.} \label{ninecellsketch}
\end{figure}

This critical path is associated to a critical displacement $u_c$
that we determine in this paper through numerical simulations of the
Ginzburg-Landau theory\cite{A57}. From the point of view of the
Ginzburg-Landau theory, pinning may be caused either by spatial
fluctuations of the critical temperature, $T_c(\vec x)$\cite{L70},
or by the mean free-path that changes the coefficient in front of
the gradient term, $\xi(\vec x)^2|({\vec \nabla} - {{2\pi
i}\over{\Phi_0}} {\vec A})\Delta|^2$. The interaction between a
vortex line and a pinning center has been considered by many authors
in the context of the Ginzburg-Landau
theory\cite{DA99,DZ02a,DZ02b,PF03}. In this paper we only consider
the no magnetic shielding limit, such that the field penetrates in
the superconductor and there is no Meissner-Ochsenfeld effect. This
situation can be viewed as a large $\kappa$ limit,
$\kappa=\lambda/\xi$ being the dimensionless Ginzburg-Landau
parameter and $\lambda$ is the penetration depth.
%
%
%

\begin{figure}[!b]
\centering
\includegraphics[width=0.5\linewidth]{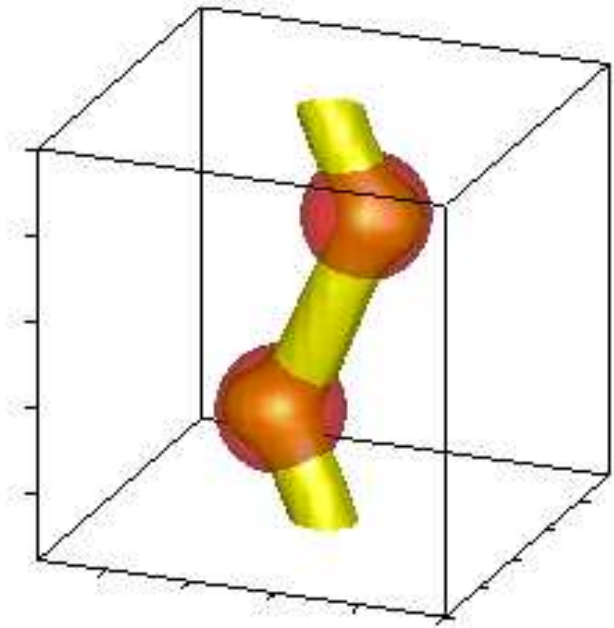}
\caption{ The vortex line remains pinned by the two displaced
insulating spheres of radius $R=1.8\xi$ in an unit cell with
$L=Lz=12\xi$. The picture is the isosurface $|\Delta|^2=0.2507$,
which means a density value approximately one fourth of its maximum
value. The pinning spheres are shown in dark gray color.}
\label{ucellvortex}
\end{figure}

The $z$-axis is the direction of the magnetic induction, and so, of
the applied magnetic field, and perpendicular to it is the x-axis.
In case of no defects, the vortex is a straight line oriented along
the $z$-axis. The defects are equally spaced along the $z$ axis and
assume alternate displacements $u$ and -$u$ along the $x$ axis.
Although the zigzag of defects fit into a single plane the problem
is genuinely three-dimensional because the defects are coherence
length size spheres. Consider the position of two defects,
$(L/2-u,L/2,Lz/4)$ and $(L/2+u,L/2,Lz/4)$. All others are obtained
from these two forming an infinite lattice in the $x$,$y$ and $z$
directions. The position of the all others defects can be obtained
by translation of $L$ along the $x$ or $y$ axis and by $Lz$ along
the $z$ axis. The periodicity due to the zigzag of defects allow to
describe this system through a orthorhombic unit cell of height
$Lz$, and of square basis $Lx=Ly=L$, with two pinning centers
inside, at the above positions $(L/2-u,L/2,Lz/4)$ and
$(L/2+u,L/2,3Lz/4)$ with respect to a coordinate frame whose origin
is at the left-down corner of the unit cell. Figures
\ref{ucellsketch} and \ref{ninecellsketch} give a pictorial view of
the geometrical arrangement of pinning centers inside the
superconductor which is under investigation here.

\begin{figure}[!t]
\centering
\includegraphics[width=0.5\linewidth]{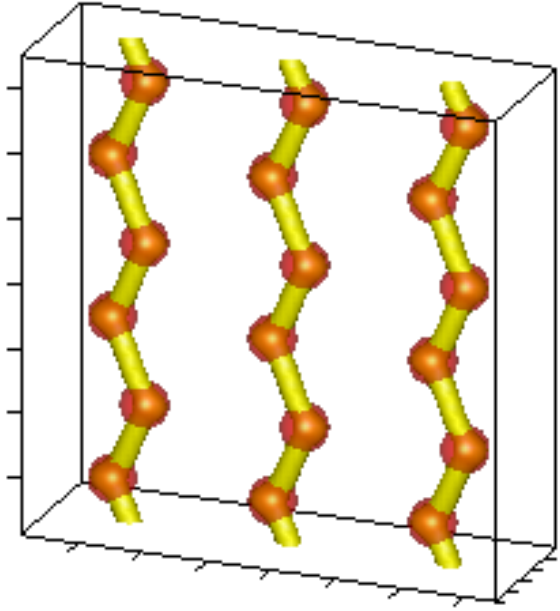}
\caption{Nine equivalent unit cells are patched here displaying the
vortex lines pinned by the spheres, as described in figure
\ref{ucellvortex}.} \label{ninecellsvortex}
\end{figure}

The numerical results obtained from simulations of the
Ginzburg-Landau theory yield the order parameter in each point of
space. These results are visualized as iso-surfaces of the density,
that is of the modulus of the order parameter squared, taken at a
fixed value, as shown in figures \ref{ucellvortex} and
\ref{ninecellsvortex}. Notice that figure \ref{ninecellsvortex}
corresponds to nine equivalent unit cells, as shown in figure
\ref{ucellvortex}, patched together. Many other possible pinned
vortex configurations are possible in the pinning center arrangement
displayed in figure \ref{ninecellsketch}. Figure
\ref{ninecellsvortex} gives one among many other, but it is the only
one described by the unit cell of figure \ref{ucellvortex}. However
these other possibilities are not considered here, although the
present method is able to treat them by searching vortex
configurations in larger unit cells containing several pairs of
pinning centers. We restrict our study here to the minimum deviation
of the vortex from the straight line, which is well described by the
present choice of unit cell. The present study involves some
parameters related to the unit cell, $L_z$ and $L$, and to the
pinning center, $u$ and $R$. The unit cell basis is fixed to
$L=12\xi$, and so our goal is to numerically obtain the Helmholtz
free energy $F(u,Lz,R)$. For each point we obtain a density plot
such as shown in figure \ref{ucellvortex}.
%
%
This paper is organized as follows.
In section \ref{Theoretical approach}, we discuss out theoretical
approach. In sections \ref{Results} we show
the results obtained through numerical simulations. In section
\ref{Conclusions} we summarize the main results of the work.

\section{Theoretical approach}
\label{Theoretical approach}

We find numerical solutions of the Ginzburg-Landau theory in the
unit cell using a mesh grid to describe it\cite{DA99,DZ02a,DZ02b}.
The distance between two consecutive mesh points, $a$, is equal to
$0.5\xi$, consistent to the fact that $a$ must be smaller than the
coherence length $\xi$, which is the minimum physical scale of the
Ginzburg-Landau theory. Thus the number of mesh points describing
the unit cell square basis is $P^2$, and so $P=25$ since $L =
a(P-1)$. Notice that the pinning sphere also imposes a limit on the
mesh parameter which cannot be larger than the pinning center
radius, that is $a=0.5\xi<R$, otherwise there will be one or no mesh
point describing the pinning region. In this case the order
parameter does not vanish inside the defect but just undergoes a
drop on its value. The energy density functional of the
Ginzburg-Landau theory\cite{A57} is expressed in units of the
critical field energy density\cite{DZ02b}, $H_c^2/4\pi$ and the
order parameter density, $\Delta|^2$ is dimensionless, varying
between 0 and 1. The concept of unit cell brings a periodicity to
the problem, and so the search for the free energy minimum must be
done under a constraint, the number of vortices inside the unit
cell, $\vec{\nu}\phi_0$, described by integers:
$\vec{\nu}=n_x\hat{x} + n_y\hat{y} + n_z\hat{z}$. The magnetic
induction, which is the average of the local field taken over the
unit cell volume, $\vec{B}=\int_v \vec{h} d^3r/V$, has a
relationship to these integers, and in reduced units is
$\vec{B}(\vec{x})=2\pi\kappa
\xi^2(n_x\hat{x}/L.L_z+n_y\hat{y}/L.L_z+n_z\hat{z}/L^2)$. In the
present paper we only consider a single vortex oriented along the
$z$-axis, hence $\vec{\nu}=\hat{z}$.
\begin{eqnarray}
F = \int {{dv}\over{V}} \;\tau(\vec x) \left[ \xi^2
\left|({\vec \nabla} - {{2\pi i}\over{\Phi_0}} {\vec A})\Delta\right|^2 -
\left|\Delta\right|^2 \right] + {1 \over 2} \left|\Delta\right|^4, \label{eq:glth}
\end{eqnarray}
The function $\tau(\vec{x})$ is a step-like function used to describe
the pinning spheres in this approach\cite{DZ02b}. Explicitly we have
$\tau(\vec{x})=\tau_1(\vec{x})\tau_2(\vec{x})$ and
\begin{eqnarray}
\label{eq:tau} \tau_i({\vec x}) = 1 - \frac{2}{1 + e^{{(|\vec x -
\vec x_i|/R)}^{N}}},
\end{eqnarray}
where $\tau_i$ is equal to $0$ inside and $1$ outside the $i$th
sphere. The above explicit representation of the $\tau$ function is
necessary for computational reasons and for accuracy we take that
$N=8$. In the limit $N\to\infty$, the function $\tau$ tends to the
well-known Heaviside function, $\tau(\vec{x})=\Theta\left (
\frac{|\vec{x}-\vec{x}_1|}{R}-1\right)\Theta\left (
\frac{|\vec{x}-\vec{x}_2|}{R}-1\right)$. The most significant
advantage of the present method, is that the free energy functional,
Eq.~\ref{eq:glth}, contains the appropriate boundaries conditions to
the problem. This removes the necessity of solving the theory in two
independent regions and later applying the Neumman boundary
conditions. Besides the present method easily applies to internal
regions of any shape, not just spherical, and finds its solution for
the given normal-superconductor interface\cite{DZ02b}.

\begin{figure}[!ht]
 \centering
 {
 \begin{minipage}[t]{0.45\textwidth}
 \includegraphics[width=\linewidth]{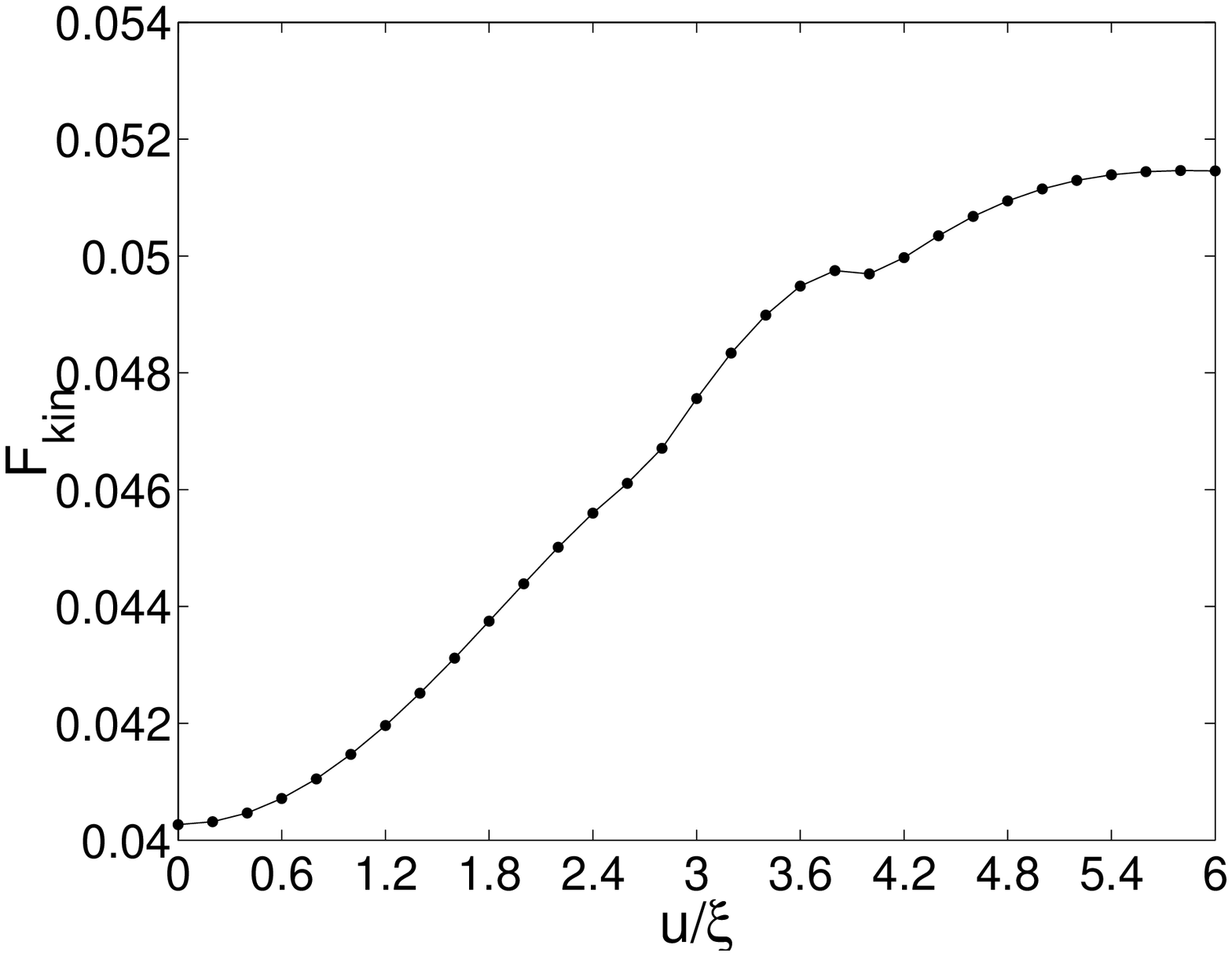}
 \end{minipage}
 }
 {
 \begin{minipage}[t]{0.45\textwidth}
 \includegraphics[width=\linewidth]{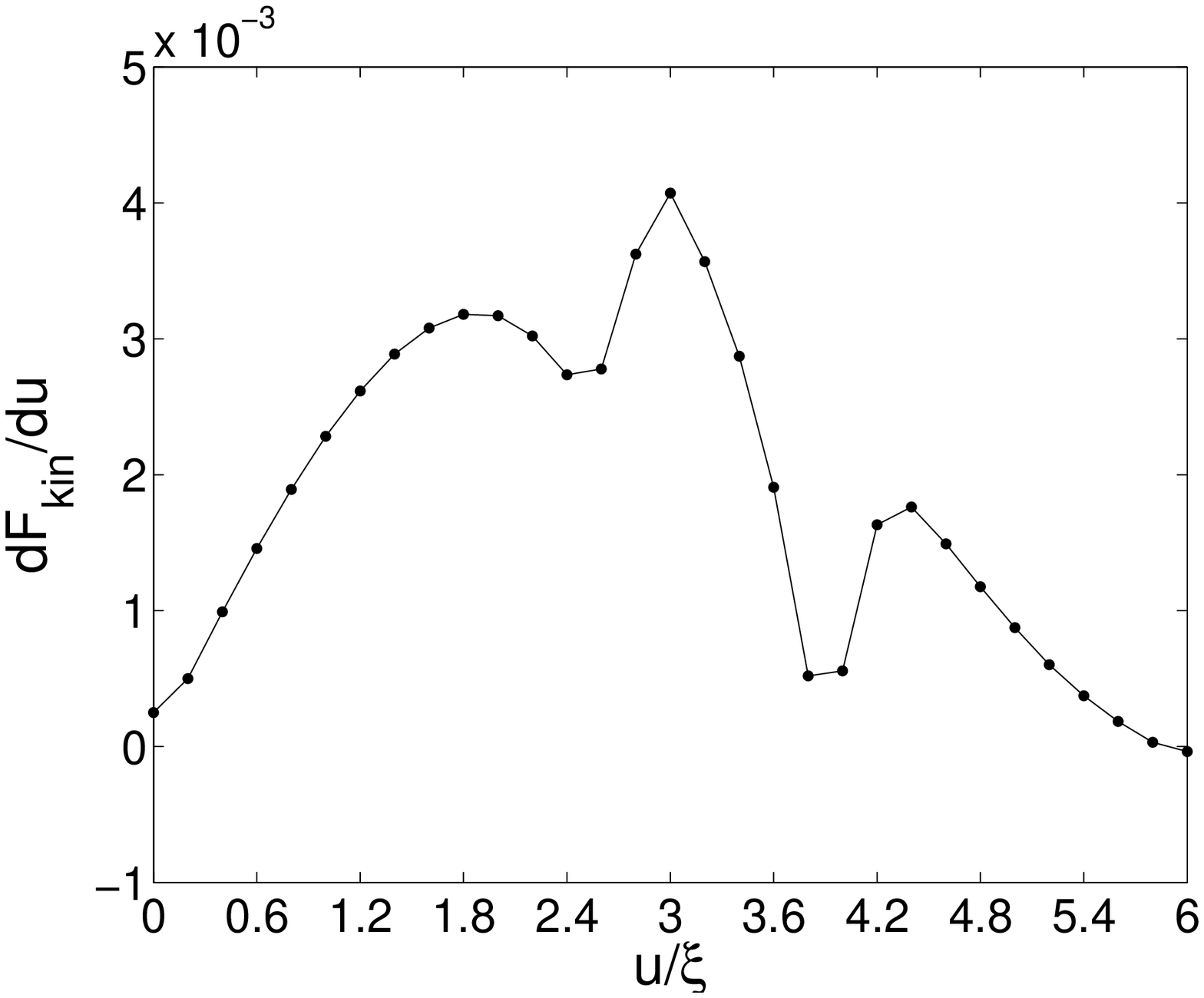}
 \end{minipage}
 }
\caption{Kinetic energy and its derivative in the left and right
part of the figure, respectively. The valleys in the derivative are
associated to depinning. The critical displacement is associated to
the minimum in these valleys. The pinning sphere radius is
$R=1.8\xi$ and $L=Lz=12\xi$} \label{fkinu1_8}
\end{figure}
%
%
\begin{figure}[!h]
 \centering
 {
 \begin{minipage}[t]{0.45\textwidth}
 \includegraphics[width=\linewidth]{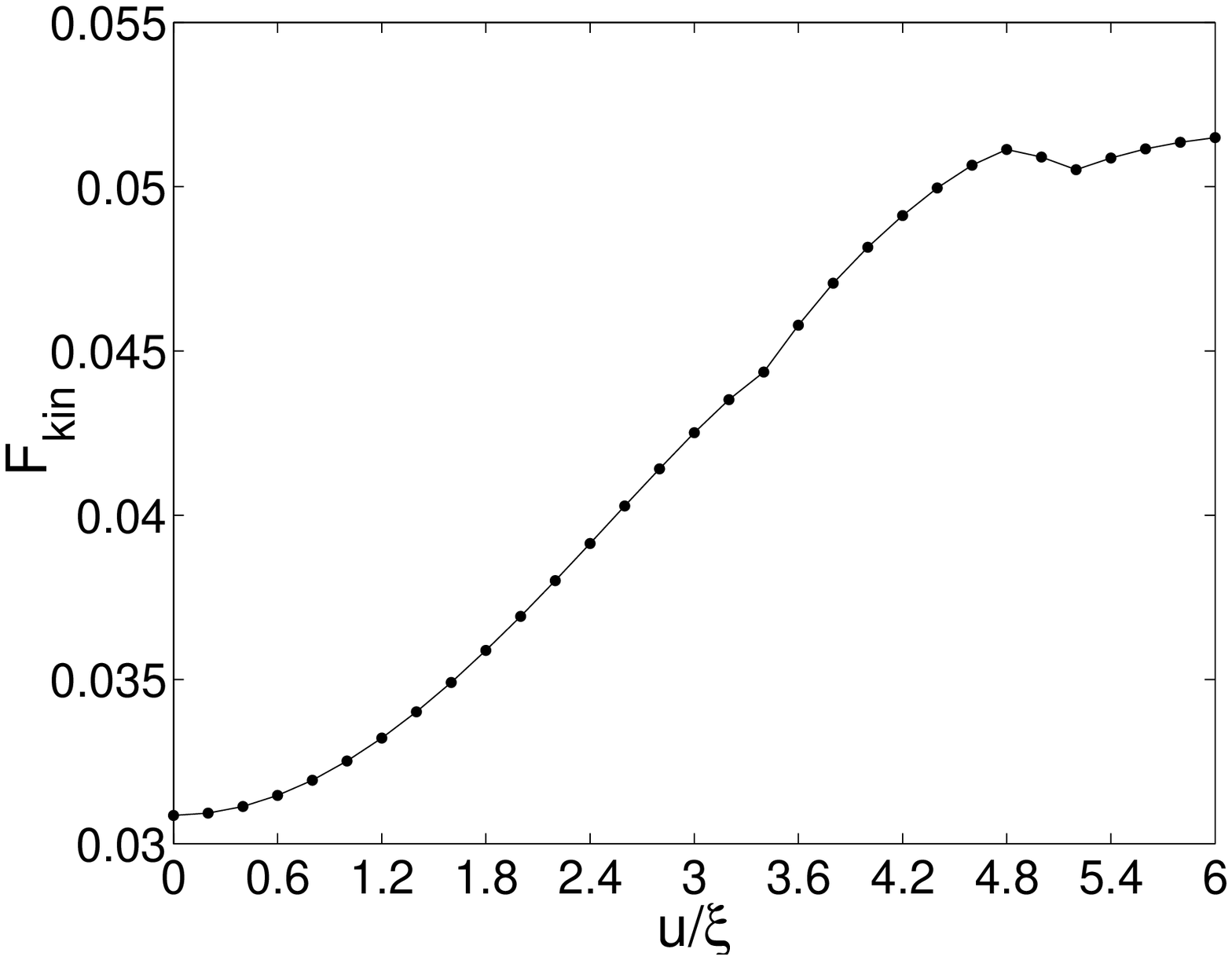}
 \end{minipage}
 }
 {
 \begin{minipage}[t]{0.45\textwidth}
 \includegraphics[width=\linewidth]{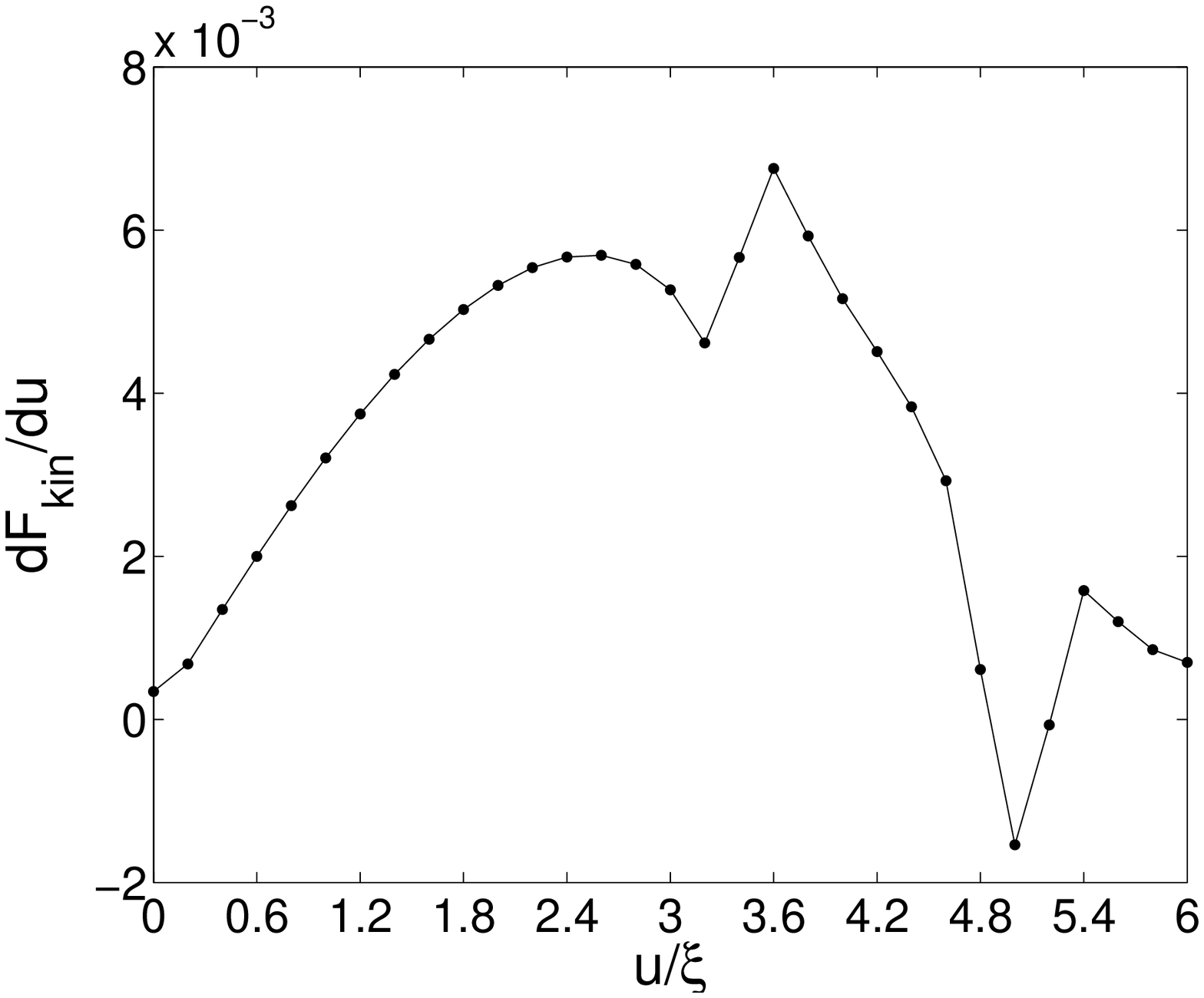}
 \end{minipage}
 }\caption{The kinetic energy and its derivative for a system with $R
= 2.4\xi$, $L=Lz=12\xi$ similar to figure \ref{fkinu1_8}. Thus the
critical displacement depends on the radius of the defect.}
\label{fkinu2_4}
\end{figure}

Since we are in the no shielding limit, the vector potential
$\vec{A}(\vec{x})$ is determined from the condition of magnetic flux
quantization inside the unit cell. The vector potential does not
participate in the minimization process of the free energy density
that only takes into account the real and imaginary parts of the
order parameter. The free energy density contains two terms. The
first term is the condensation energy density, $- \tau(\vec
x)\left|\Delta\right|^2+ {1 \over 2} \left|\Delta\right|^4$, which
in case of no vortices ($\left|\Delta\right|^2=1$) and no pinning
sphere ($\tau=1$) has the value -0.5. The presence of a pinning
sphere raises the energy since inside it the density vanishes
($\left|\Delta\right|^2=0$). And the second term, the kinetic energy
density, $\tau(\vec x)\xi^2 \left|({\vec \nabla} - {{2\pi
i}\over{\Phi_0}} {\vec A})\Delta \right|^2$. Notice that there is
kinetic energy in case of no vortices but with a defect. At the
insulating-superconducting interface $\tau$ changes from 1 to 0 and
this causes a bending of the order parameter, which has some kinetic
energy cost.
\section{Results}
\label{Results}

The critical displacement $u_c$ is better observed in the derivative
of the kinetic energy density\cite{RD04}. Figures \ref{fkinu1_8} and
\ref{fkinu2_4} shows the curves $dF_{kin}/du$ versus $u$ for two
selected radii, namely $R=1.8\xi$, and $R=2.4\xi$, respectively, and
in both cases a double hump structure with a local minimum between
them is seen. For growing displacement $u$ the first minimum of the
kinetic energy density occurs for no displacement and the second
minimum corresponds to the depinning transition $u_c$.
After the second hump there is a third minimum
associated to the depinning from the second sphere. Thus
detachment of the vortex from a pinning sphere makes the kinetic
energy density reach a minimum. To understand this consider
the local maximum that precedes the detachment. For
small $u$ the vortex core is superposed to the two pinning centers.
As long as the vortex is pinned by the two spheres, and follows the
zigzag path, the order parameter has to adjust around one single
common interface that involves both the vortex core and the pinning
centers. The maximum stretch of the vortex line is the longest
zigzag path, a configuration that demands the maximum amount of
circulating current around the vortex. Thus the maximum stretch of
the vortex line is achieved when the kinetic energy reaches its
maximum.
\begin{figure}[!b]
    \centering
    {
    \begin{minipage}[t]{0.31\textwidth}
    \centering
    \includegraphics[width=0.6\linewidth]{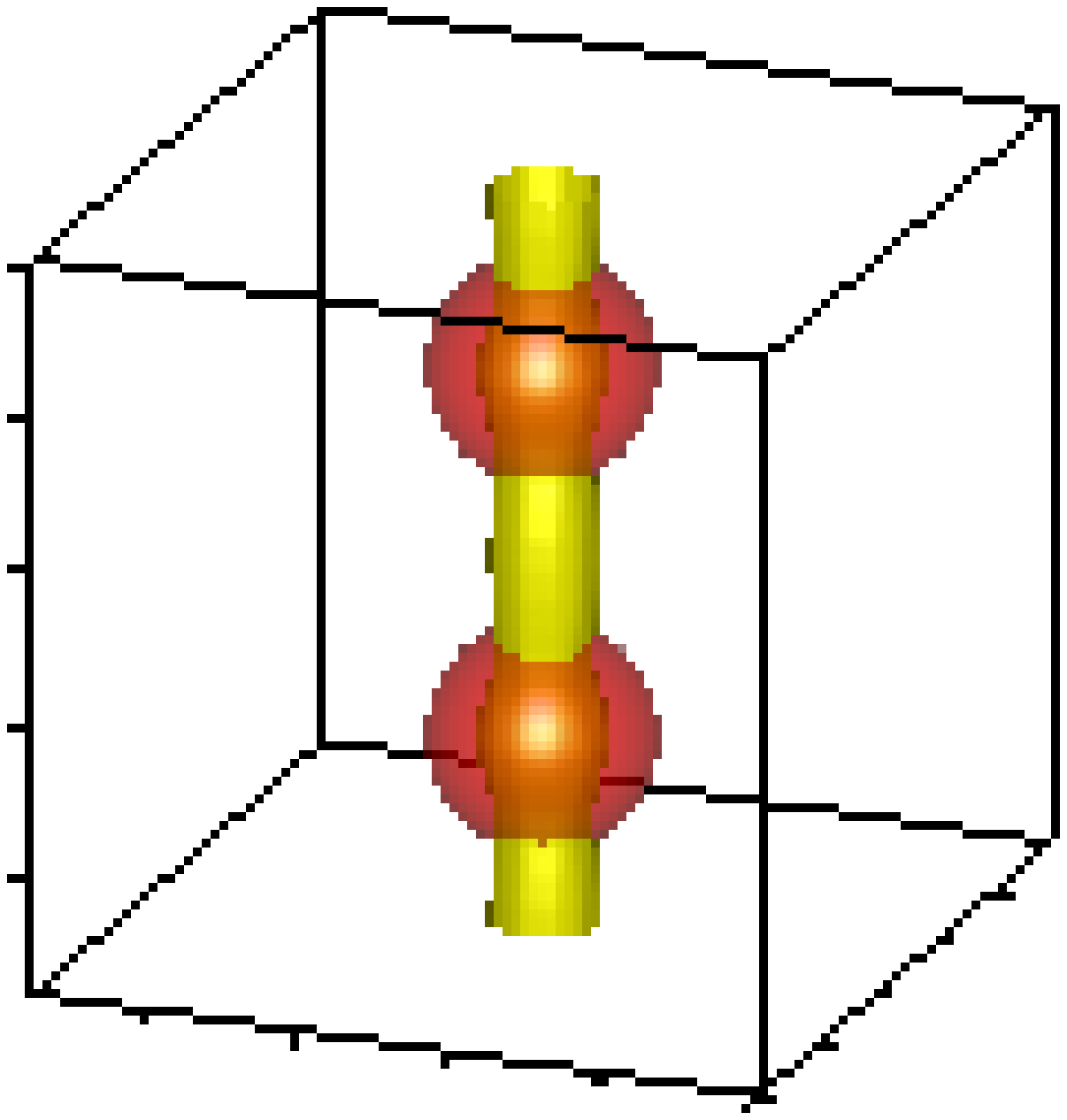}
    \end{minipage}
    }
    {
    \begin{minipage}[t]{0.31\textwidth}
    \centering
    \includegraphics[width=0.6\linewidth]{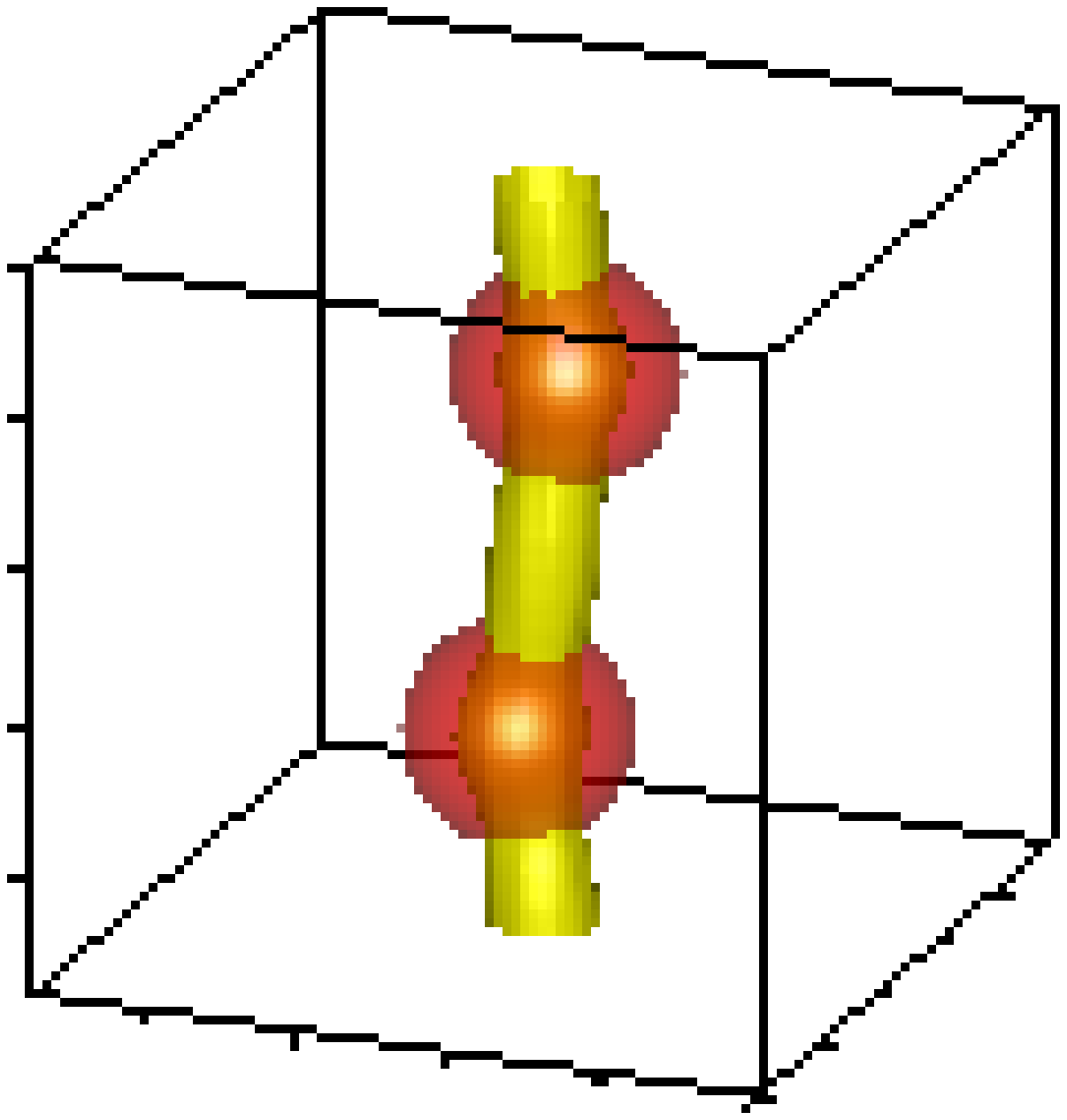}
    \end{minipage}
    }
    {
    \begin{minipage}[t]{0.31\textwidth}
    \centering
    \includegraphics[width=0.6\linewidth]{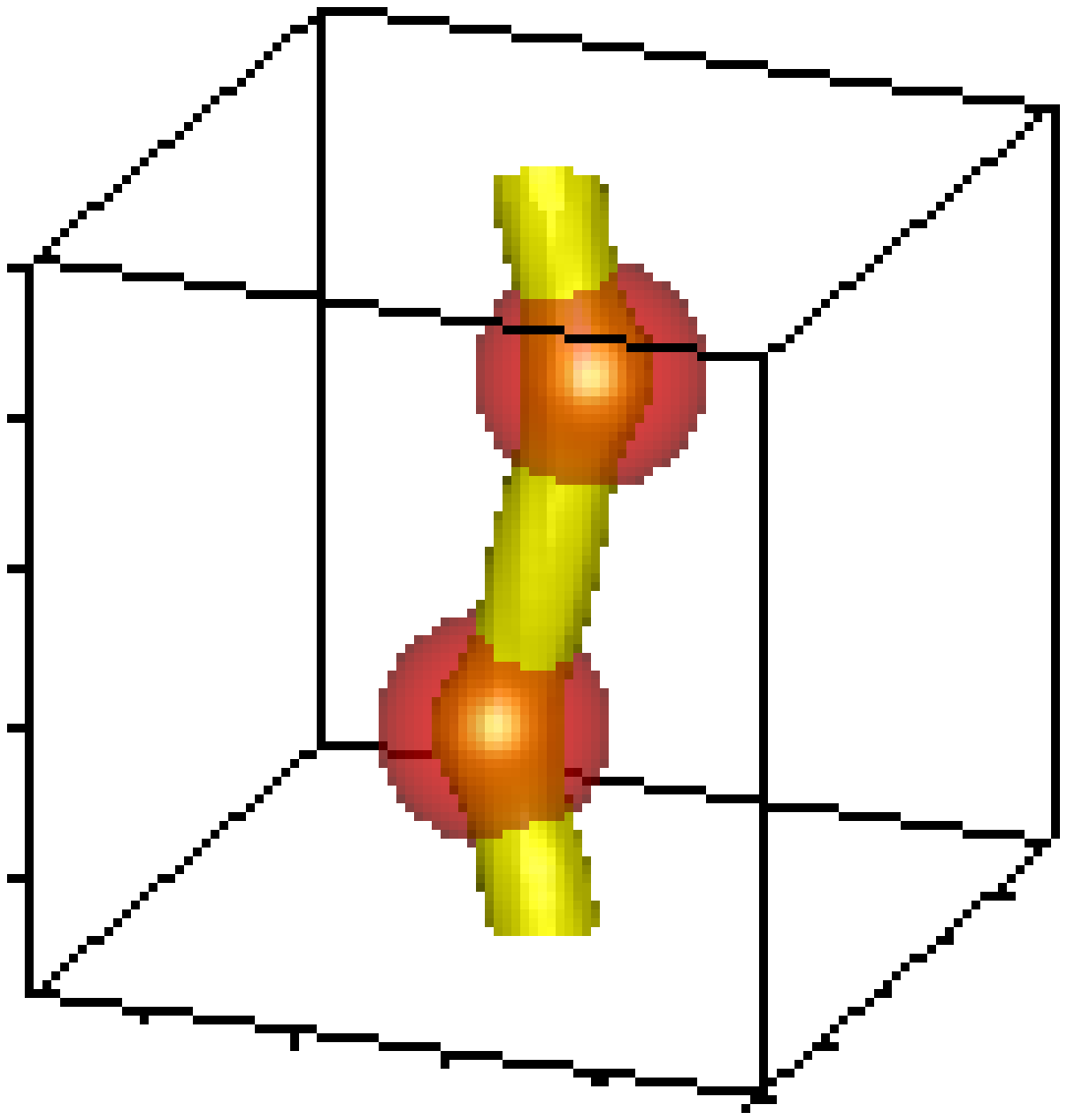}
    \end{minipage}
    }
    {
    \begin{minipage}[t]{0.31\textwidth}
    \centering
    \includegraphics[width=0.6\linewidth]{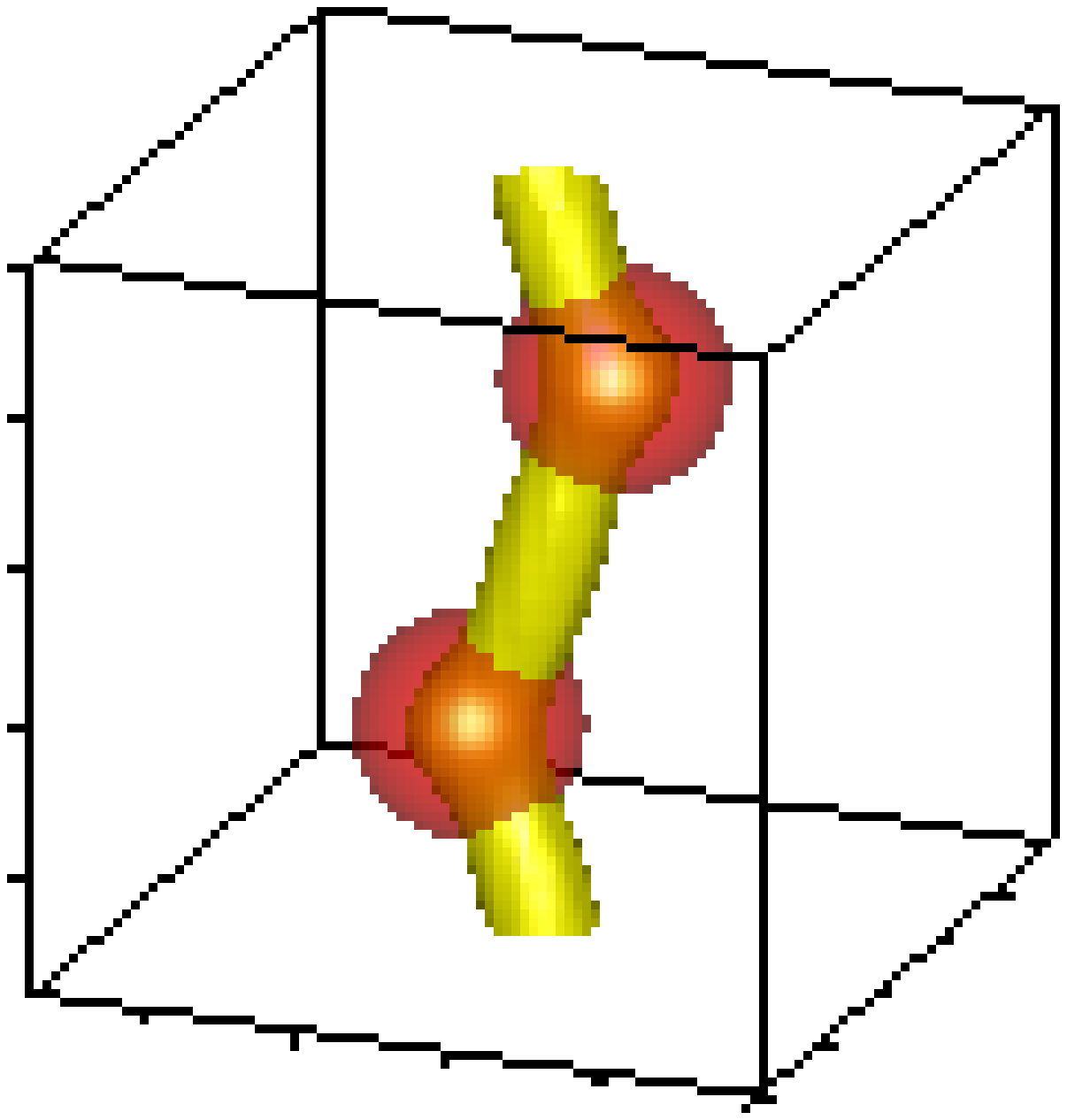}
    \end{minipage}
    }
    {
    \begin{minipage}[t]{0.31\textwidth}
    \centering
    \includegraphics[width=0.6\linewidth]{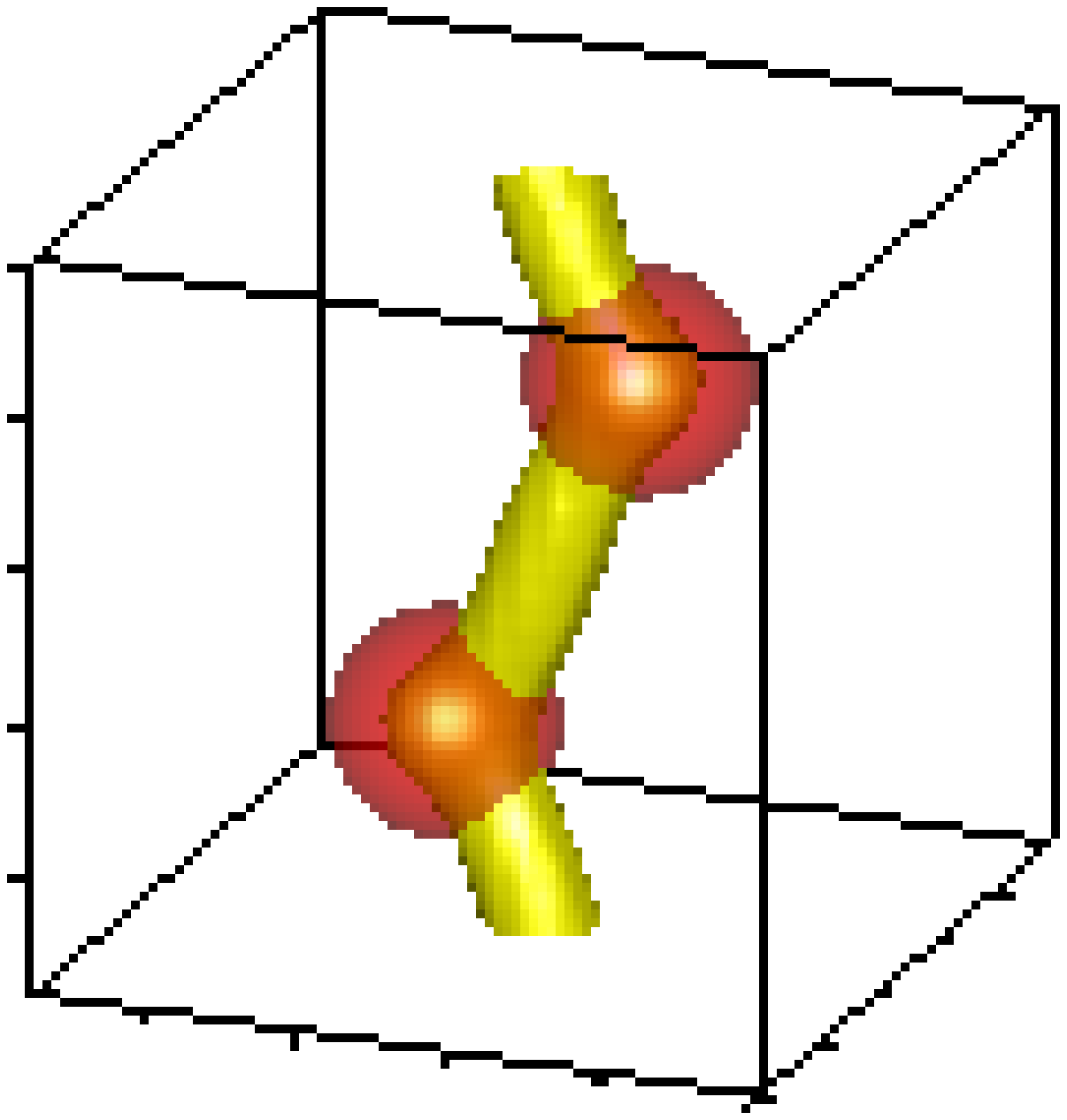}
    \end{minipage}
    }
    {
    \begin{minipage}[t]{0.31\textwidth}
    \centering
    \includegraphics[width=0.6\linewidth]{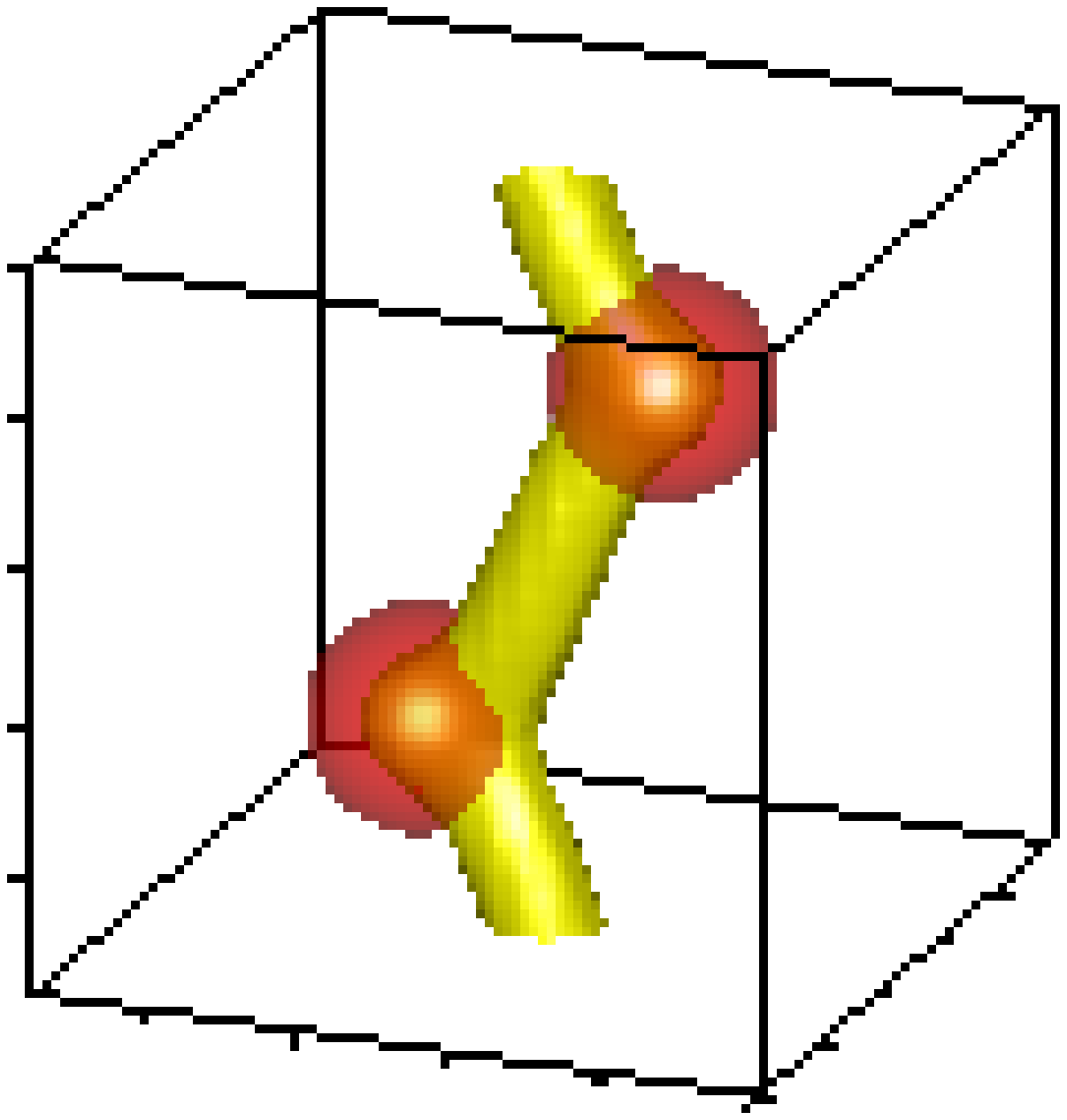}
    \end{minipage}
    }
    {
    \begin{minipage}[t]{0.31\textwidth}
    \centering
    \includegraphics[width=0.6\linewidth]{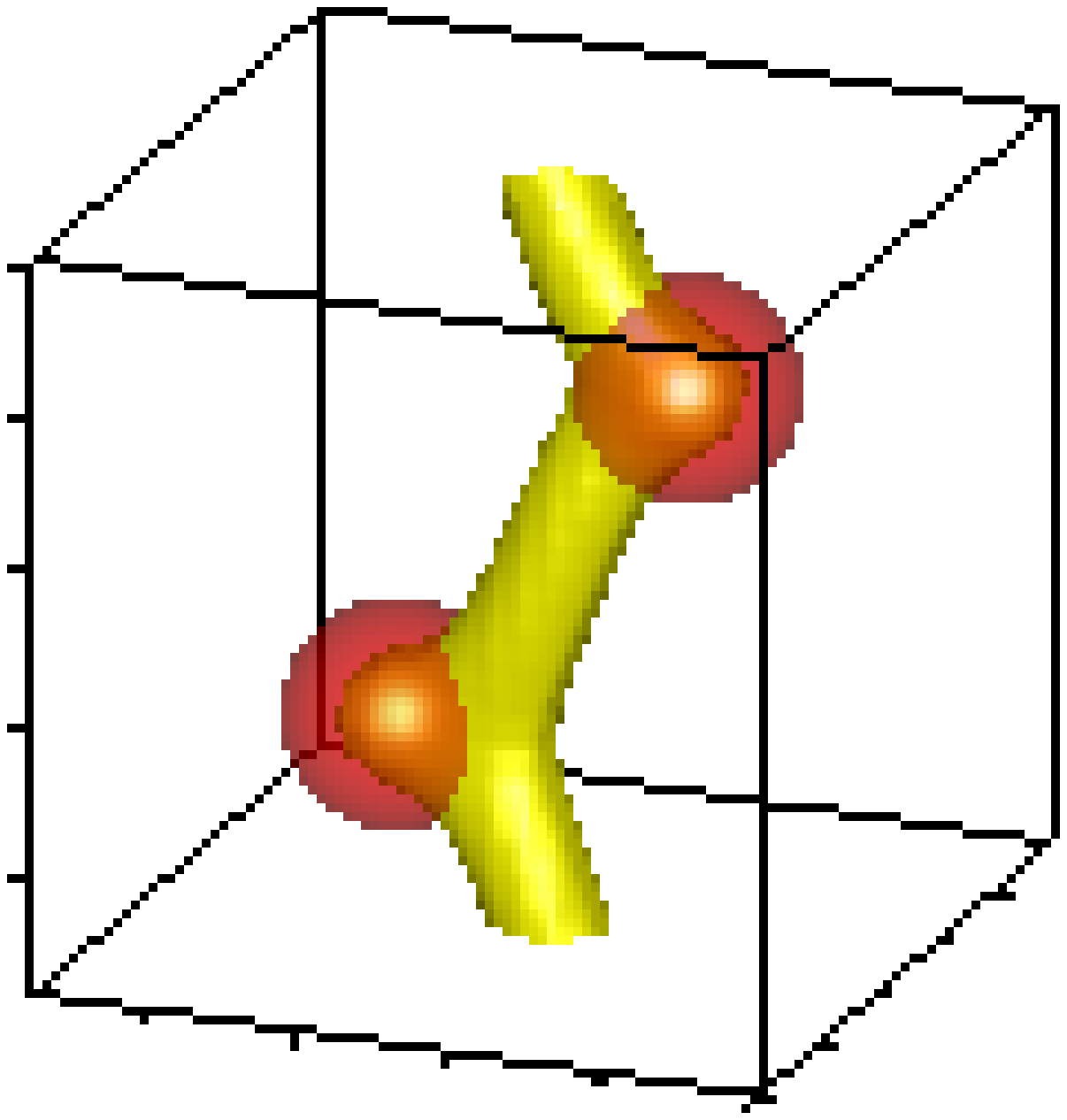}
    \end{minipage}
    }
    {
    \begin{minipage}[t]{0.31\textwidth}
    \centering
    \includegraphics[width=0.6\linewidth]{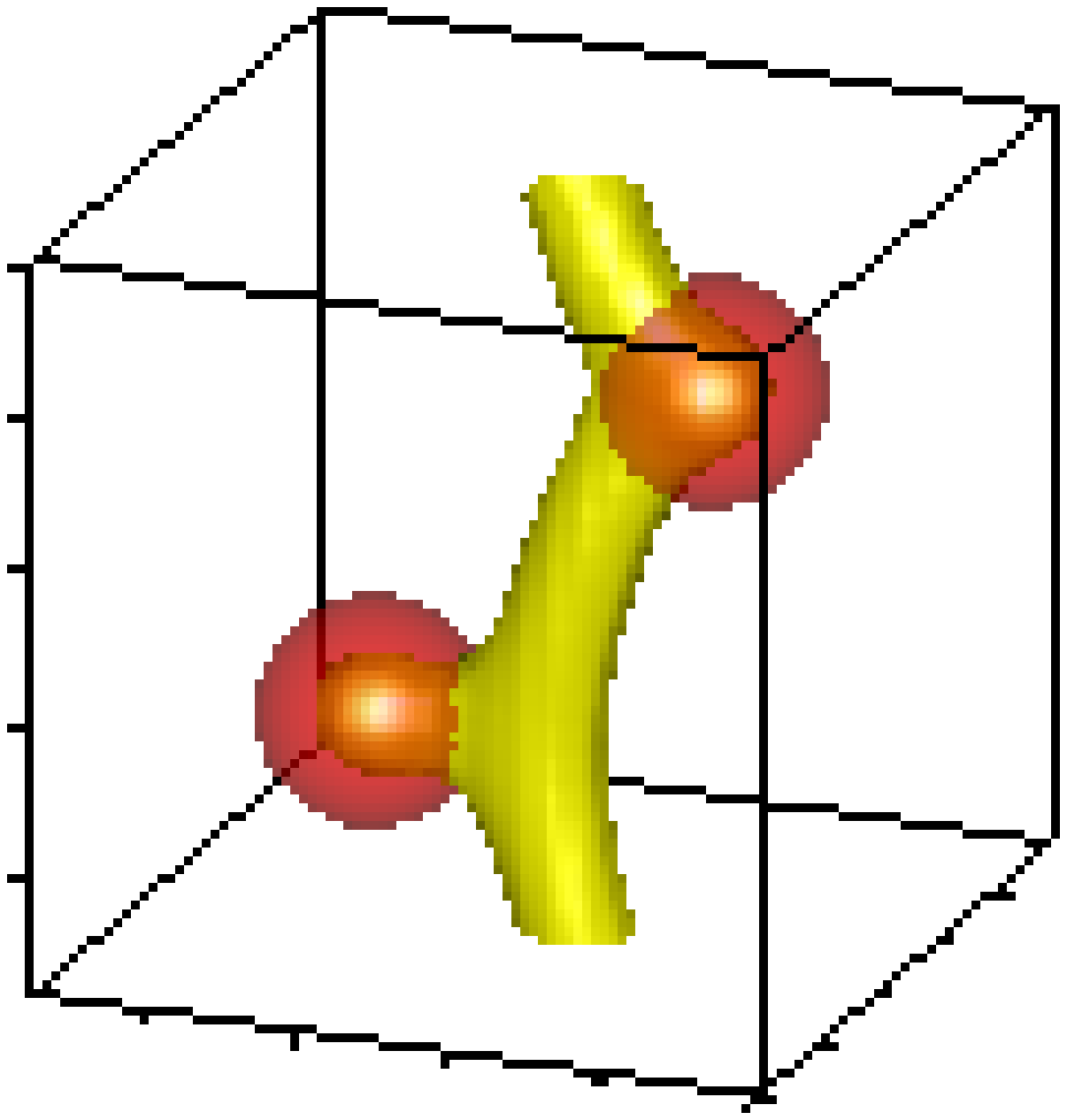}
    \end{minipage}
    }
    {
    \begin{minipage}[t]{0.31\textwidth}
    \centering
    \includegraphics[width=0.6\linewidth]{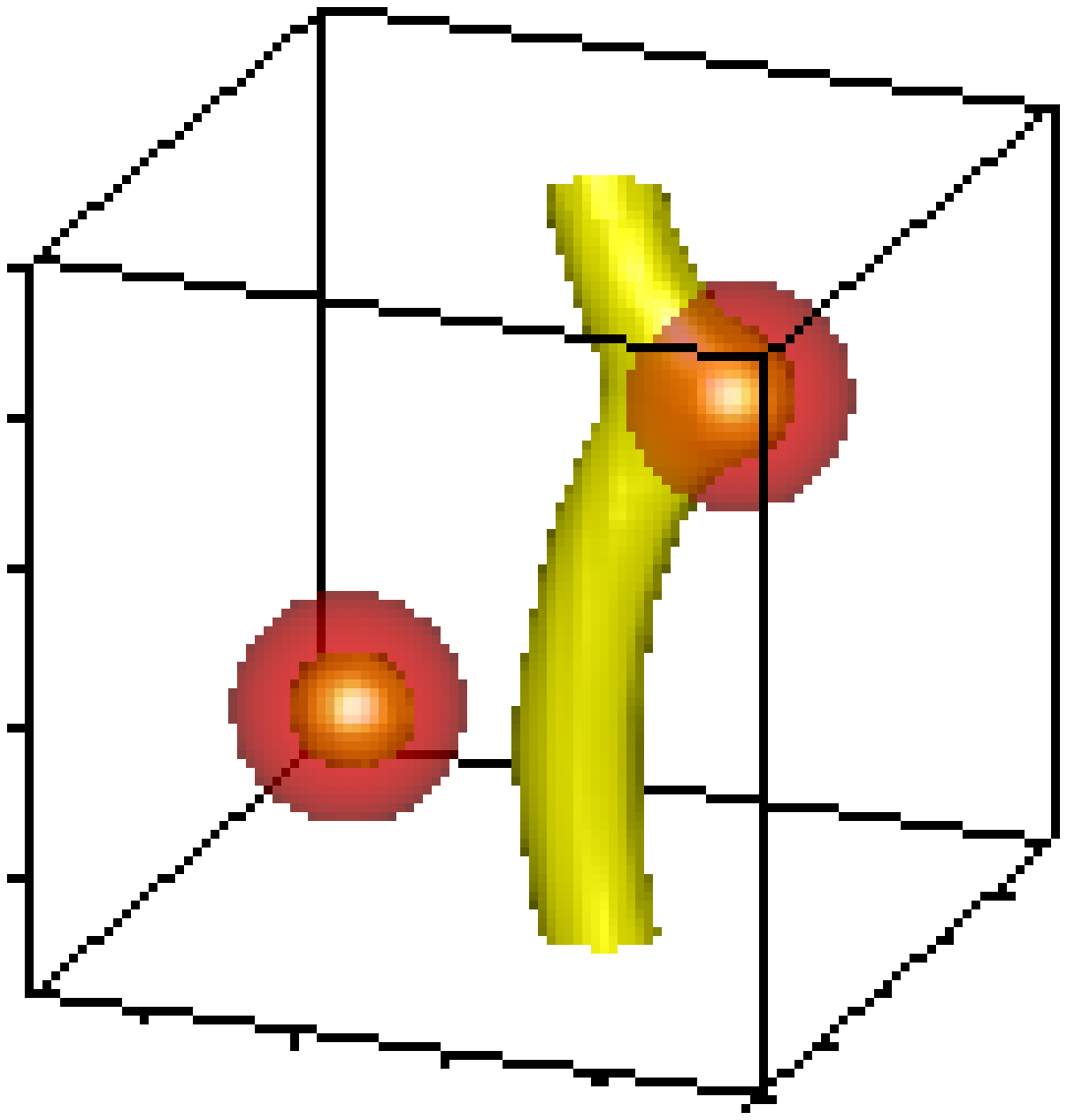}
    \end{minipage}
    }

\caption{Sequence of views of the vortex line for several growing
displacements $u/\xi$ equal to $0.0$, $0.4$, $0.8$, $1.2$, $1.6$,
$2.0$, $2.4$, $2.8$, and $3.2$. The unit cell is cubic, $L_z=L=12\xi$,
and the pinning center is $R=1.8\xi$. The isosurfaces are taken at
a density approximately equal for all cases, namely, one fourth of its maximum value, $|\Delta|^2 \approx 0.25$.}
\label{disloc}
\end{figure}
\begin{figure}[!t]
    \centering
    {
    \begin{minipage}[t]{0.31\textwidth}
    \centering
    \includegraphics[width=0.6\linewidth]{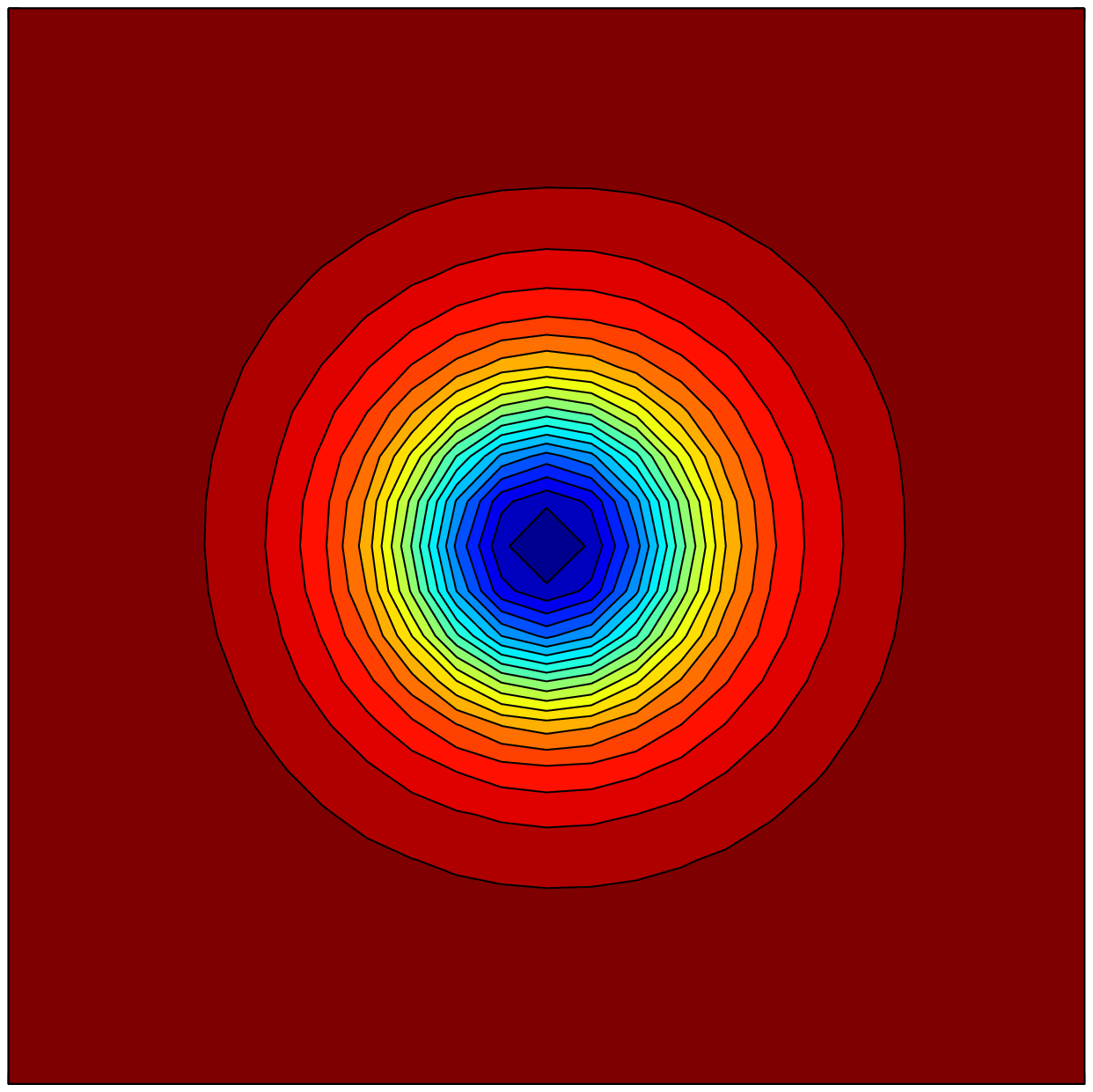}

    \end{minipage}
    }
    {
    \begin{minipage}[t]{0.31\textwidth}
    \centering
    \includegraphics[width=0.6\linewidth]{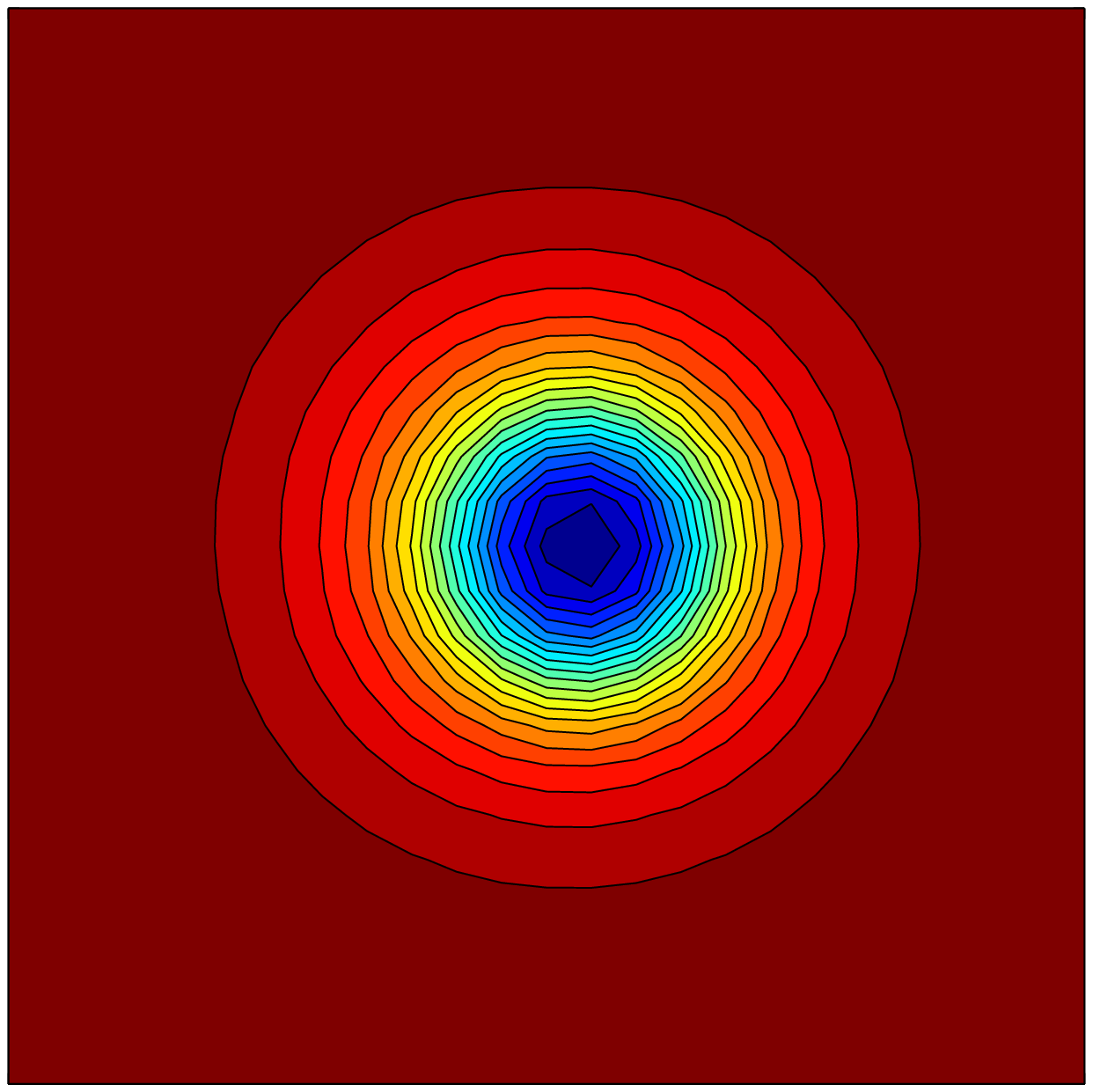}
    \end{minipage}
    }
    {
    \begin{minipage}[t]{0.31\textwidth}
    \centering
    \includegraphics[width=0.6\linewidth]{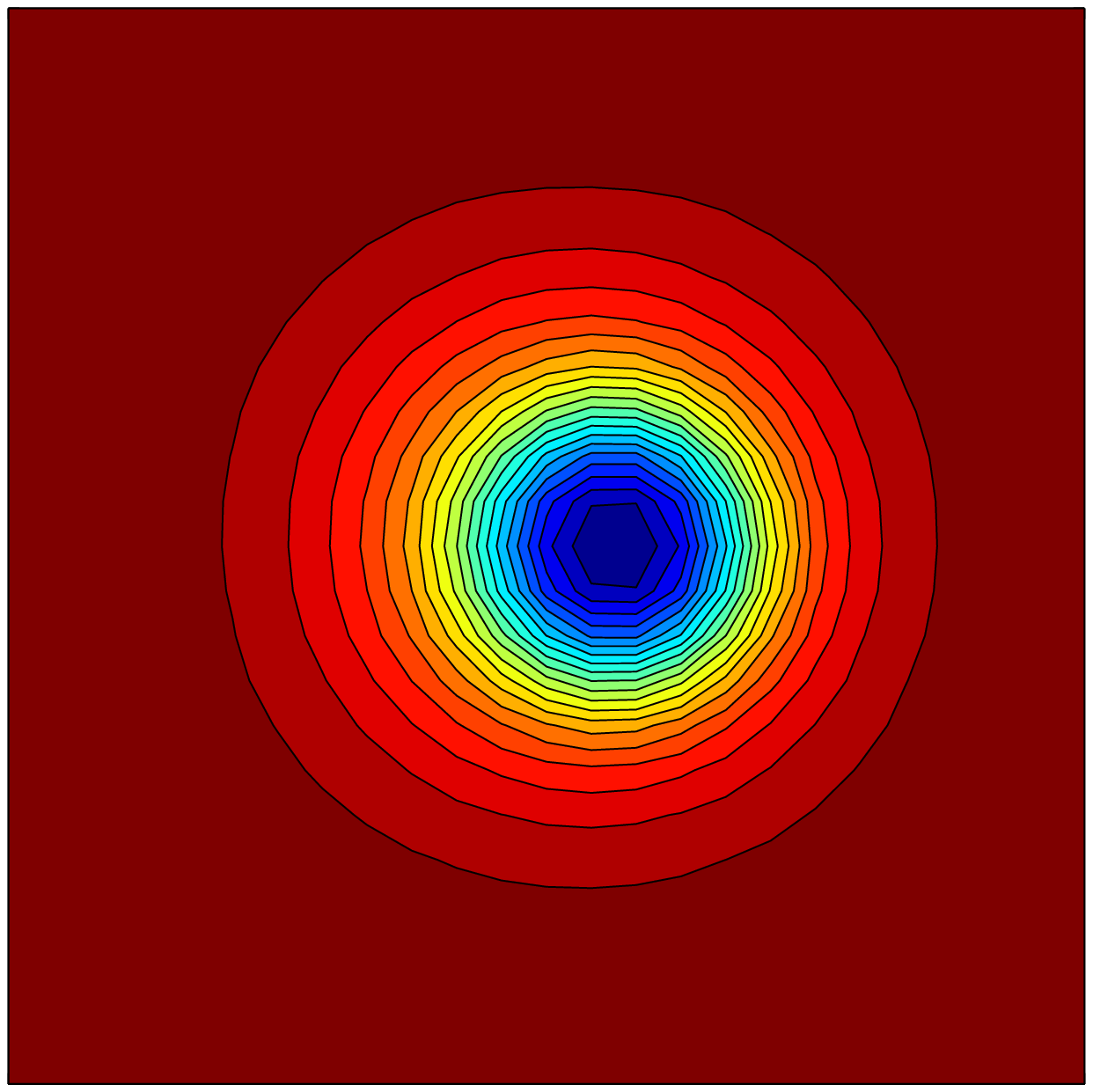}
    \end{minipage}
    }
    {
    \begin{minipage}[t]{0.31\textwidth}
    \centering
    \includegraphics[width=0.6\linewidth]{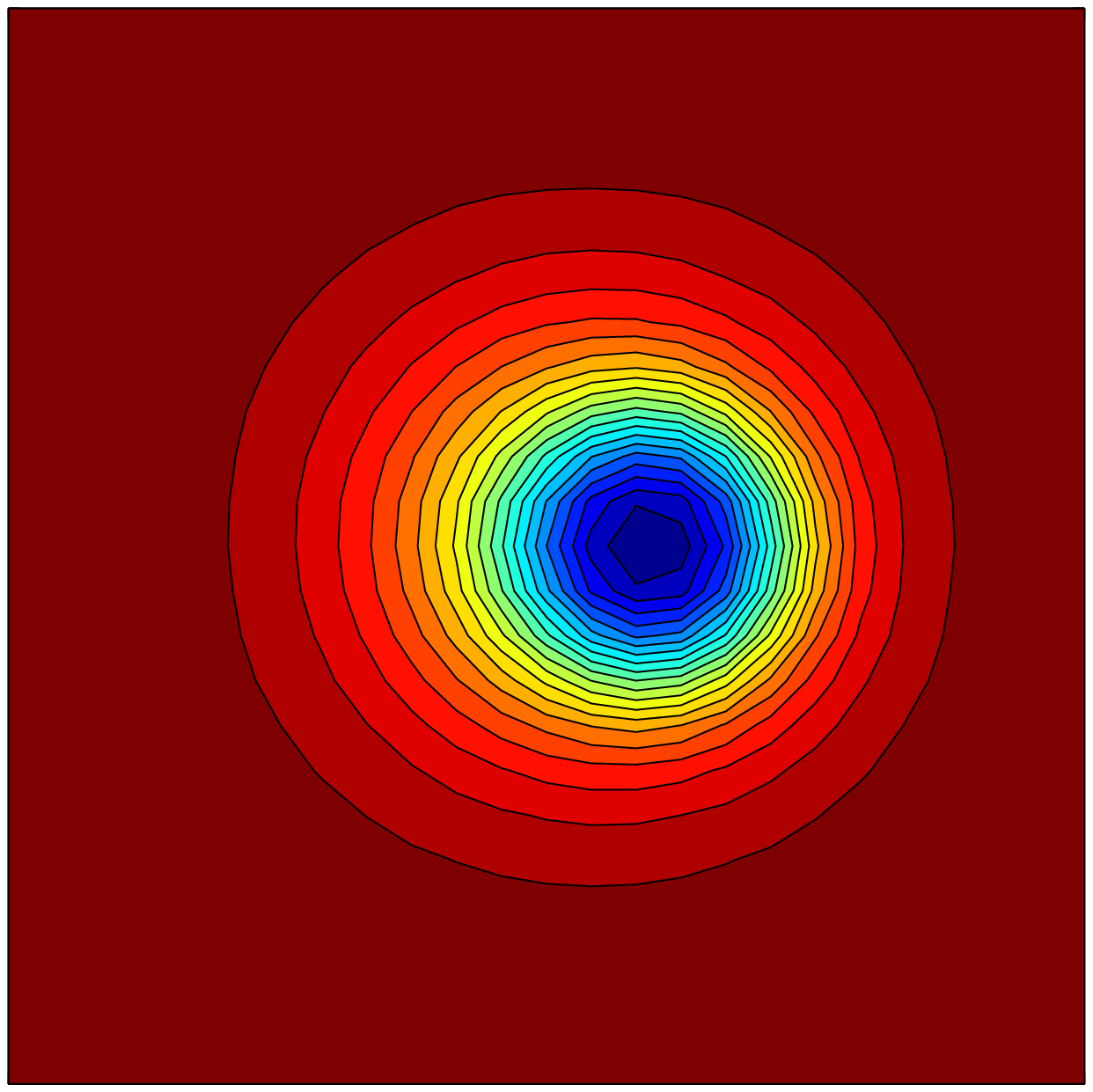}
    \end{minipage}
    }
    {
    \begin{minipage}[t]{0.31\textwidth}
    \centering
    \includegraphics[width=0.6\linewidth]{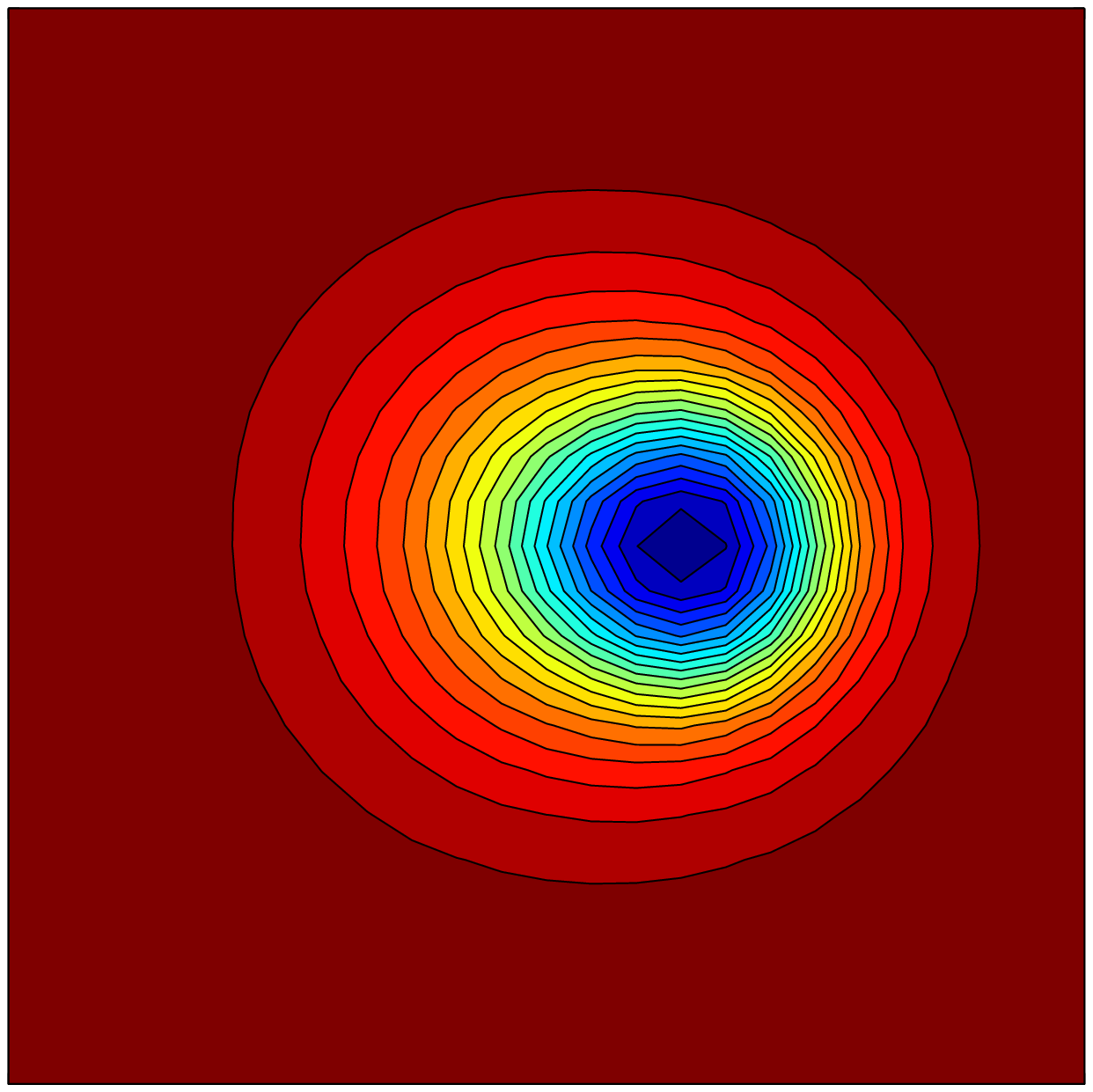}
    \end{minipage}
    }
    {
    \begin{minipage}[t]{0.31\textwidth}
    \centering
    \includegraphics[width=0.6\linewidth]{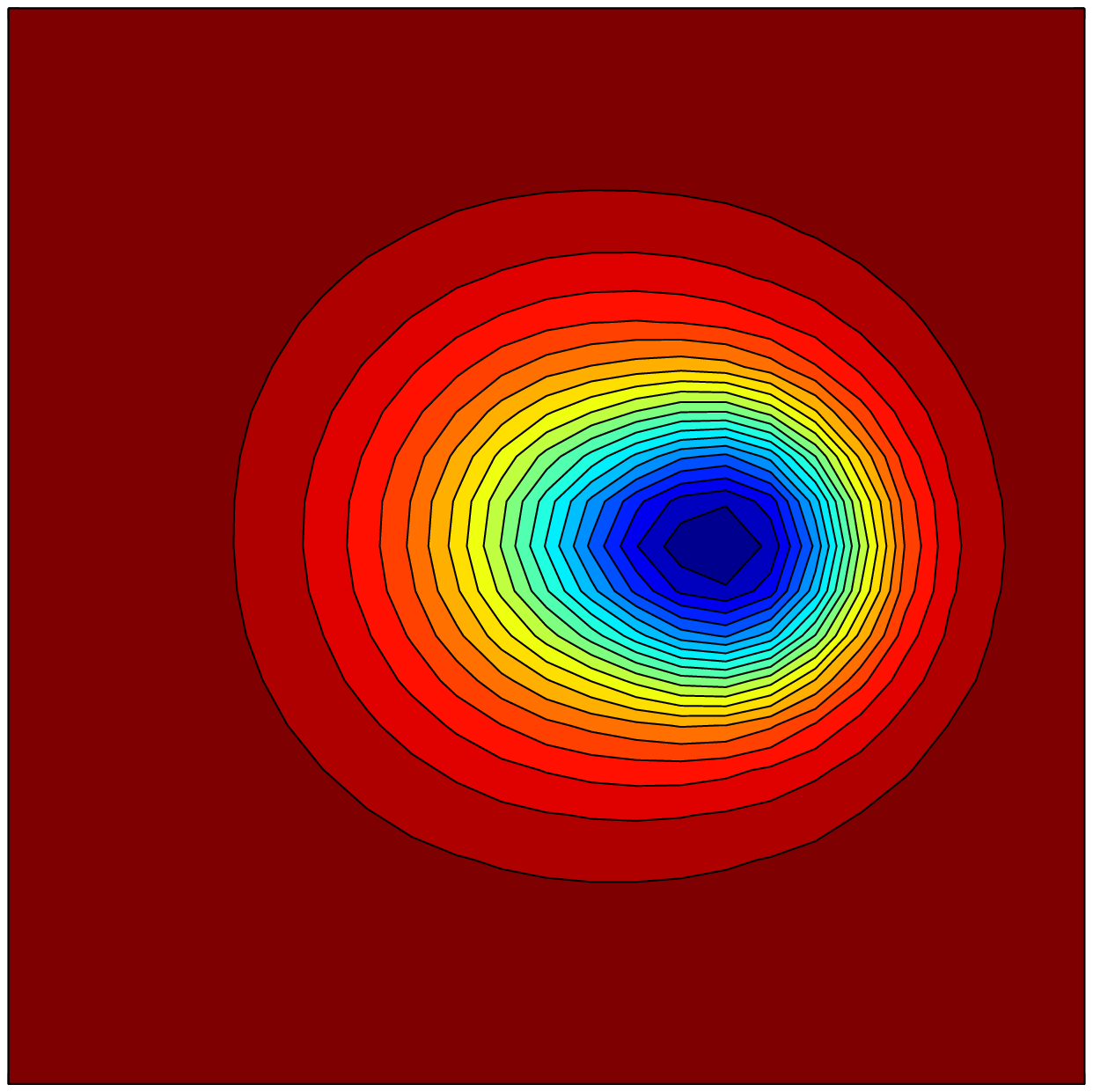}
    \end{minipage}
    }
    {
    \begin{minipage}[t]{0.31\textwidth}
    \centering
    \includegraphics[width=0.6\linewidth]{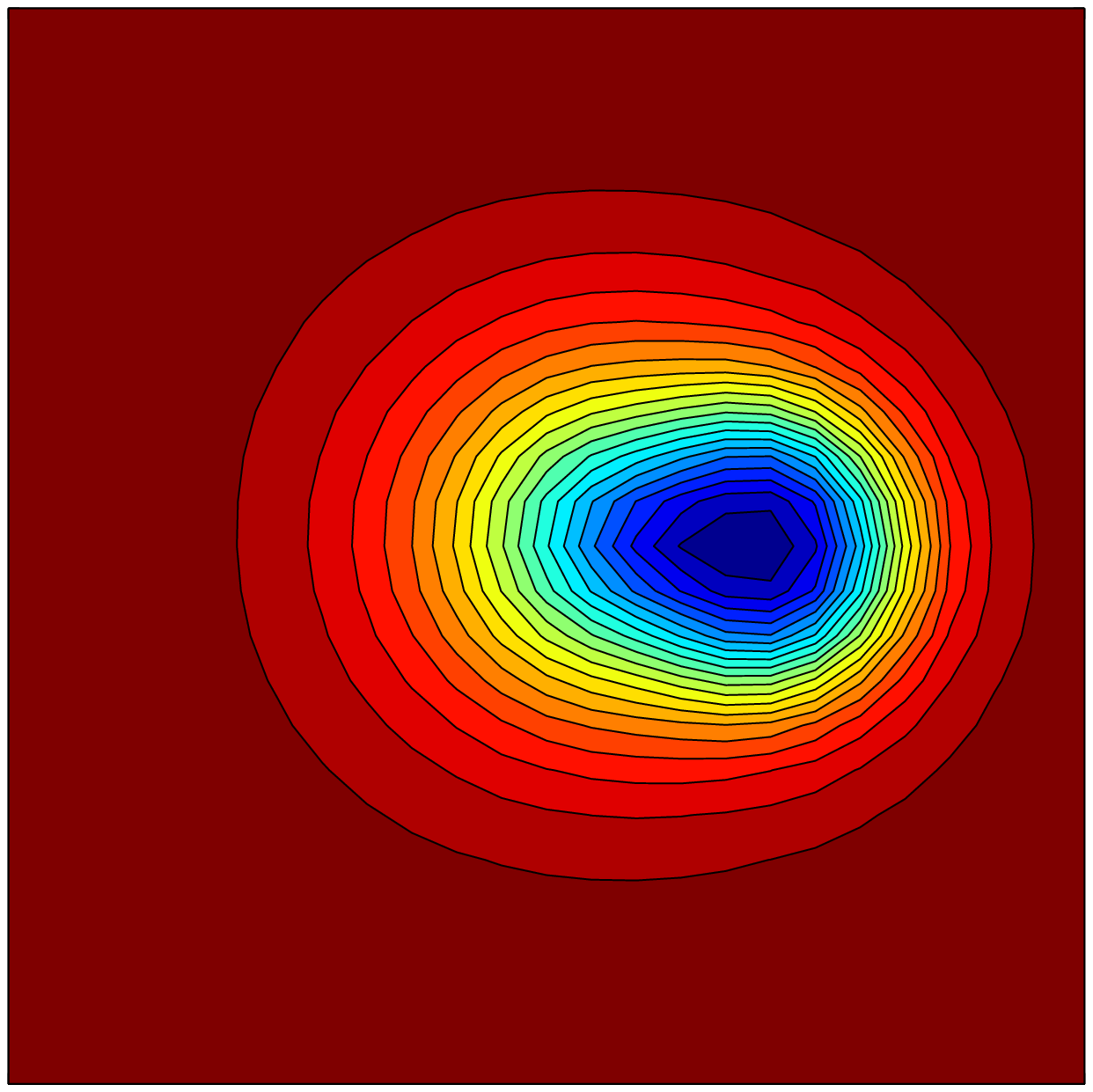}
    \end{minipage}
    }
    {
    \begin{minipage}[t]{0.31\textwidth}
    \centering
    \includegraphics[width=0.6\linewidth]{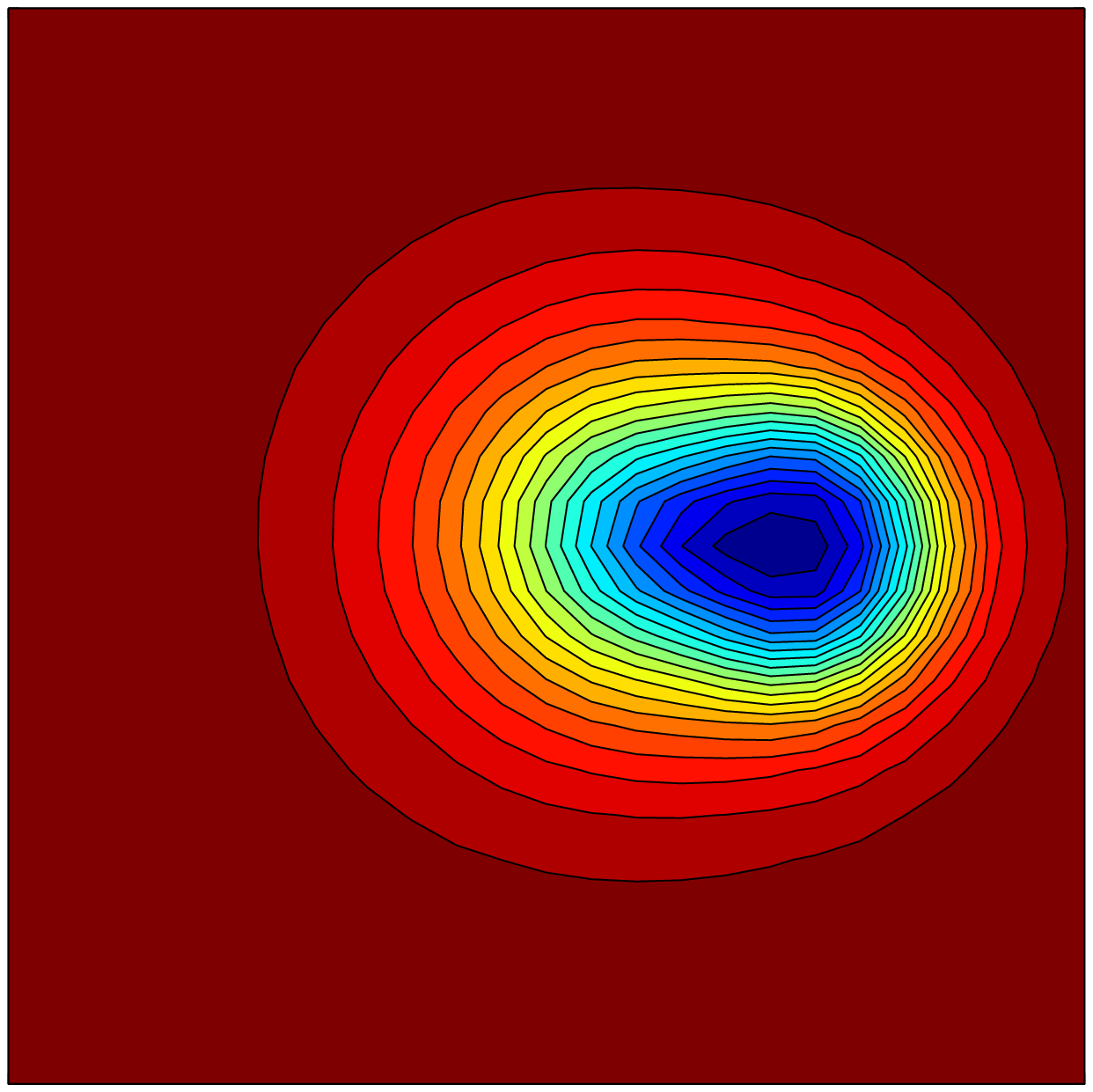}
    \end{minipage}
    }
    {
    \begin{minipage}[t]{0.31\textwidth}
    \centering
    \includegraphics[width=0.6\linewidth]{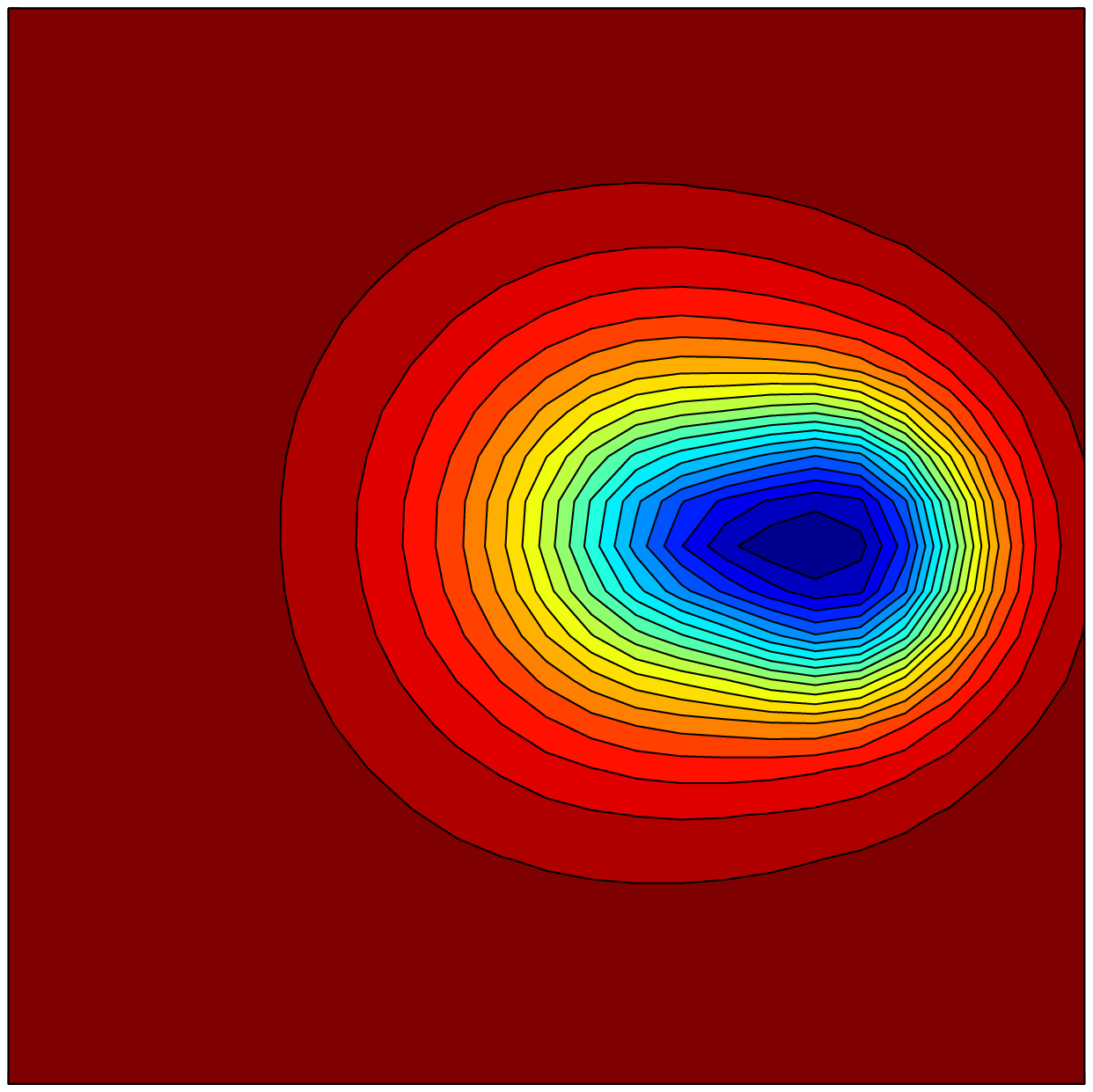}
    \end{minipage}
    }

\caption{The same sequence of views of figure \ref{disloc} is shown
here through $z=3L_z/4$ cuts of that unit cell ($L=Lz=12\xi$,$R=1.8\xi$).
For the displacement $u=3.2\xi$, the vortex
line is still pinned by upper the sphere.} \label{dislocutz2}
\end{figure}
\begin{figure}[!h]
   \centering
    {
    \begin{minipage}[t]{0.31\textwidth}
    \centering
    \includegraphics[width=0.6\linewidth]{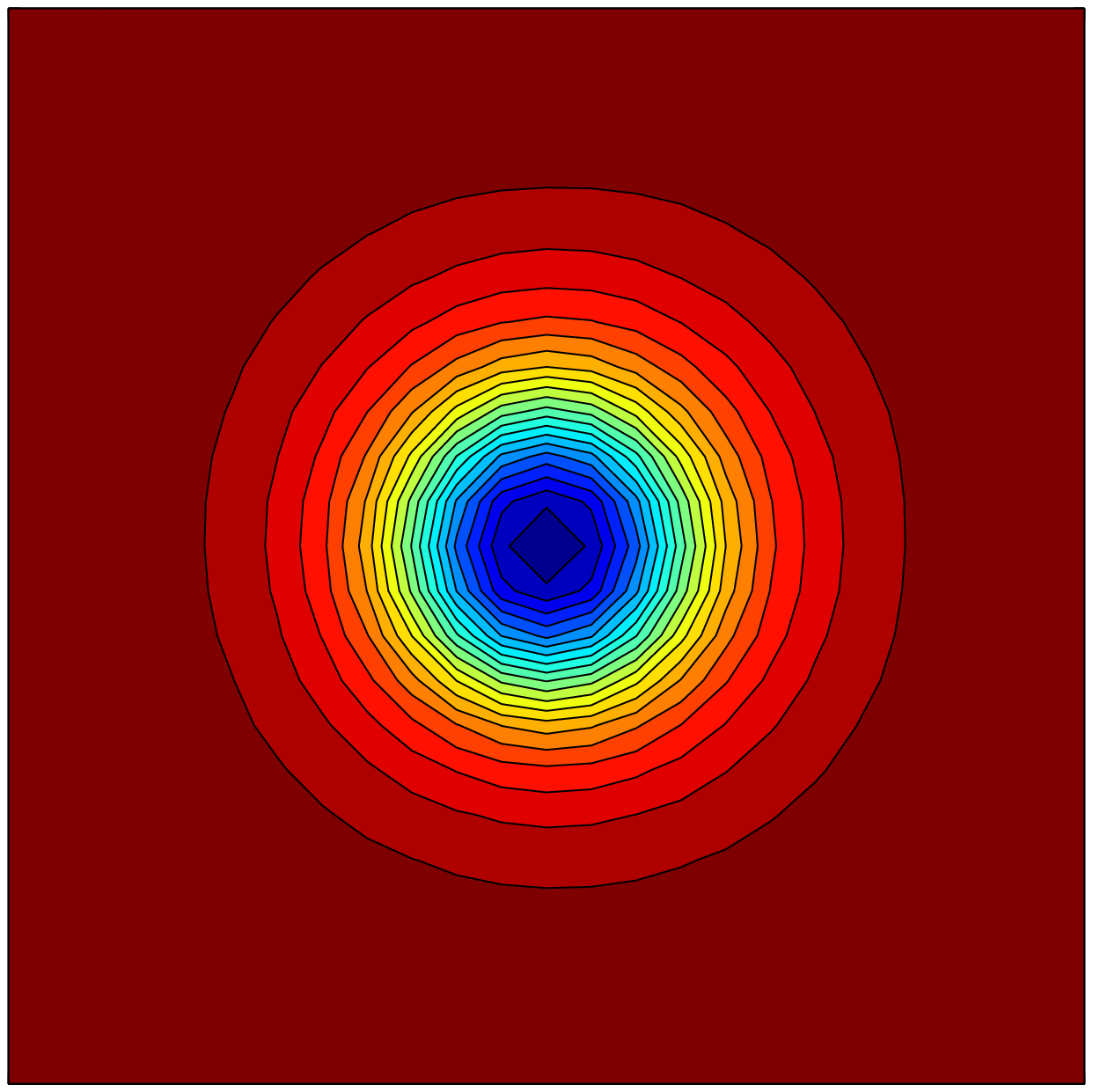}
    \end{minipage}
    }
    {
    \begin{minipage}[t]{0.31\textwidth}
    \centering
    \includegraphics[width=0.6\linewidth]{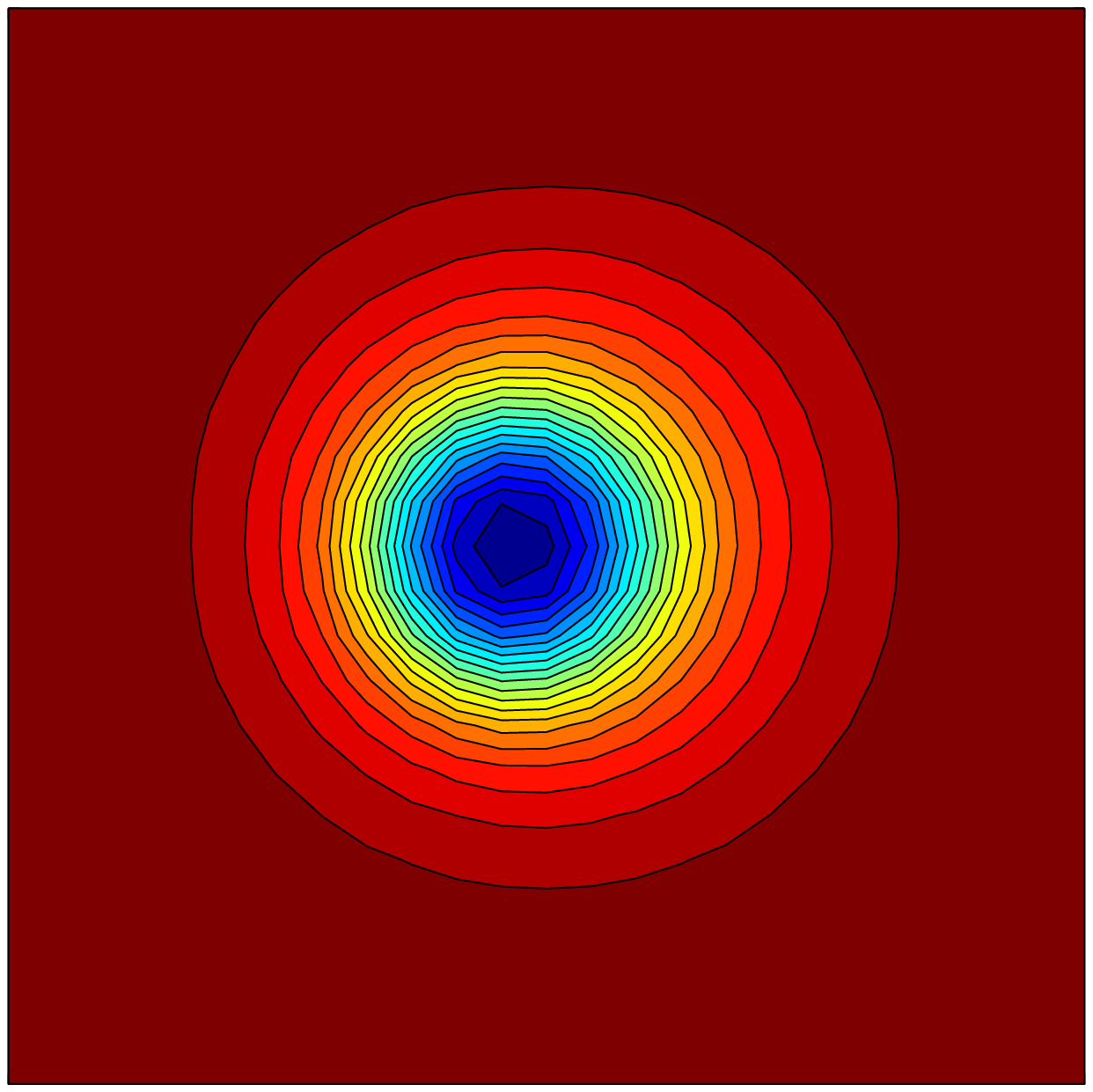}
    \end{minipage}
    }
    {
    \begin{minipage}[t]{0.31\textwidth}
    \centering
    \includegraphics[width=0.6\linewidth]{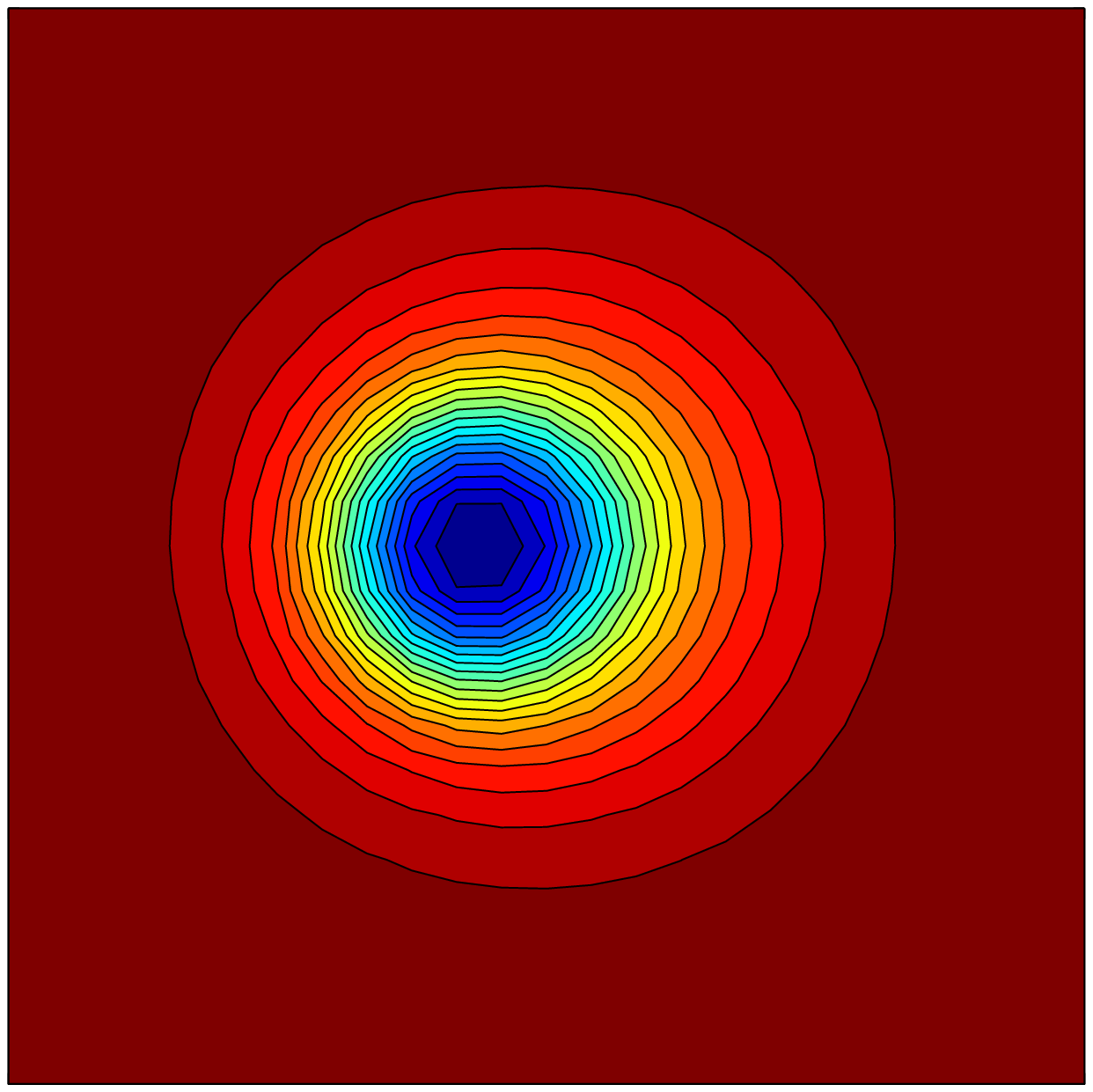}
    \end{minipage}
    }
    {
    \begin{minipage}[t]{0.31\textwidth}
    \centering
    \includegraphics[width=0.6\linewidth]{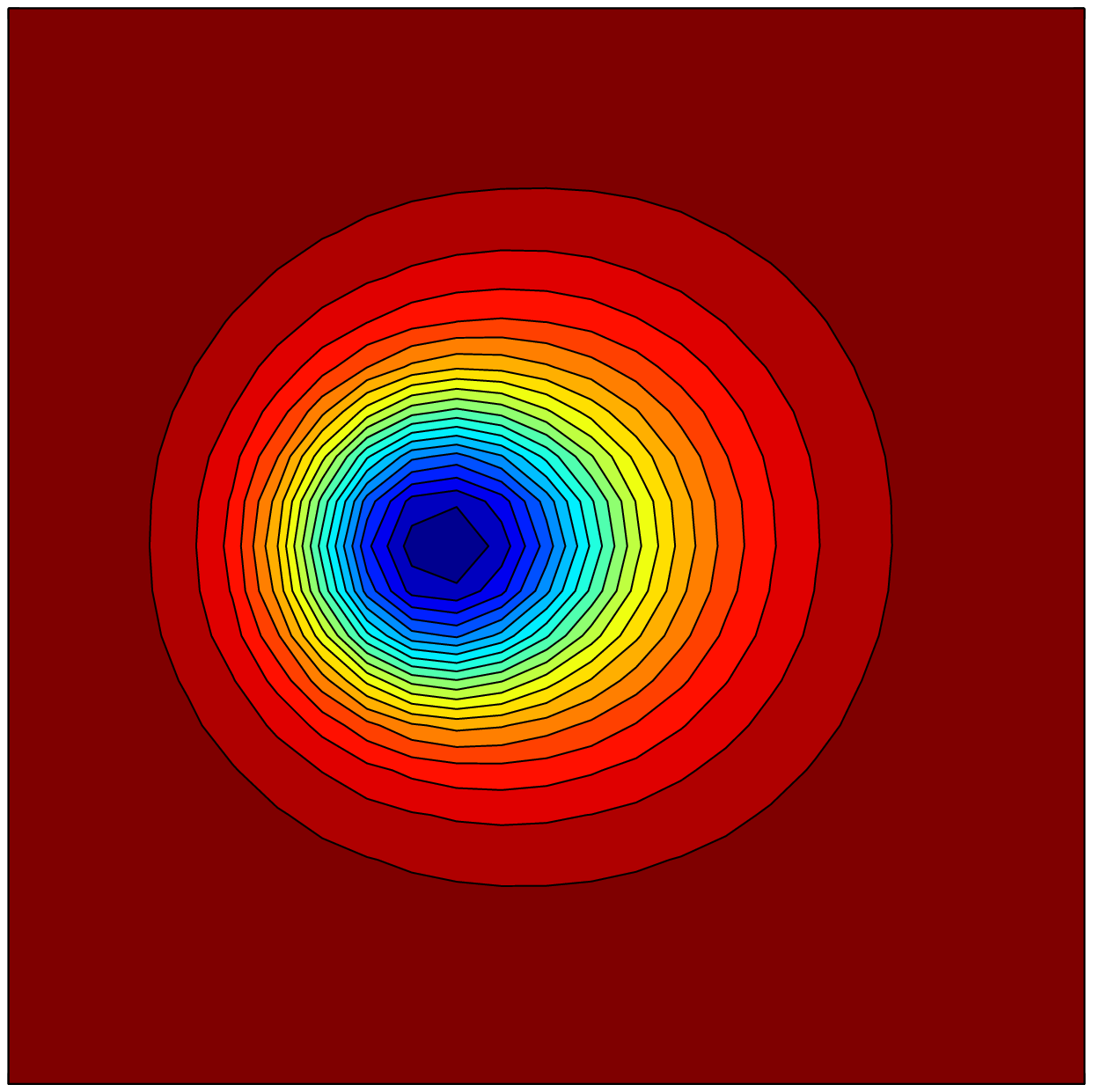}
    \end{minipage}
    }
    {
    \begin{minipage}[t]{0.31\textwidth}
    \centering
    \includegraphics[width=0.6\linewidth]{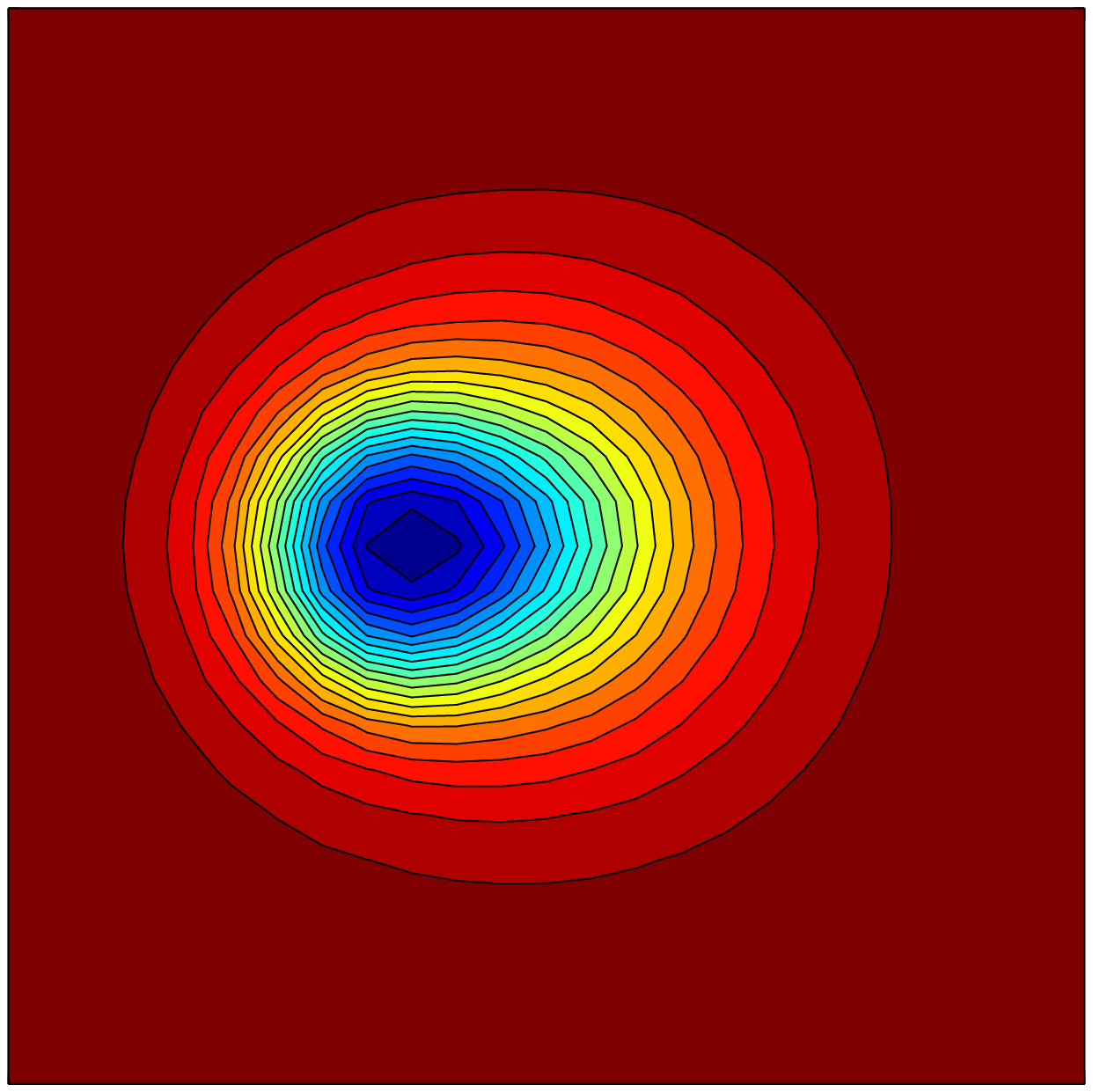}
    \end{minipage}
    }
    {
    \begin{minipage}[t]{0.31\textwidth}
    \centering
    \includegraphics[width=0.6\linewidth]{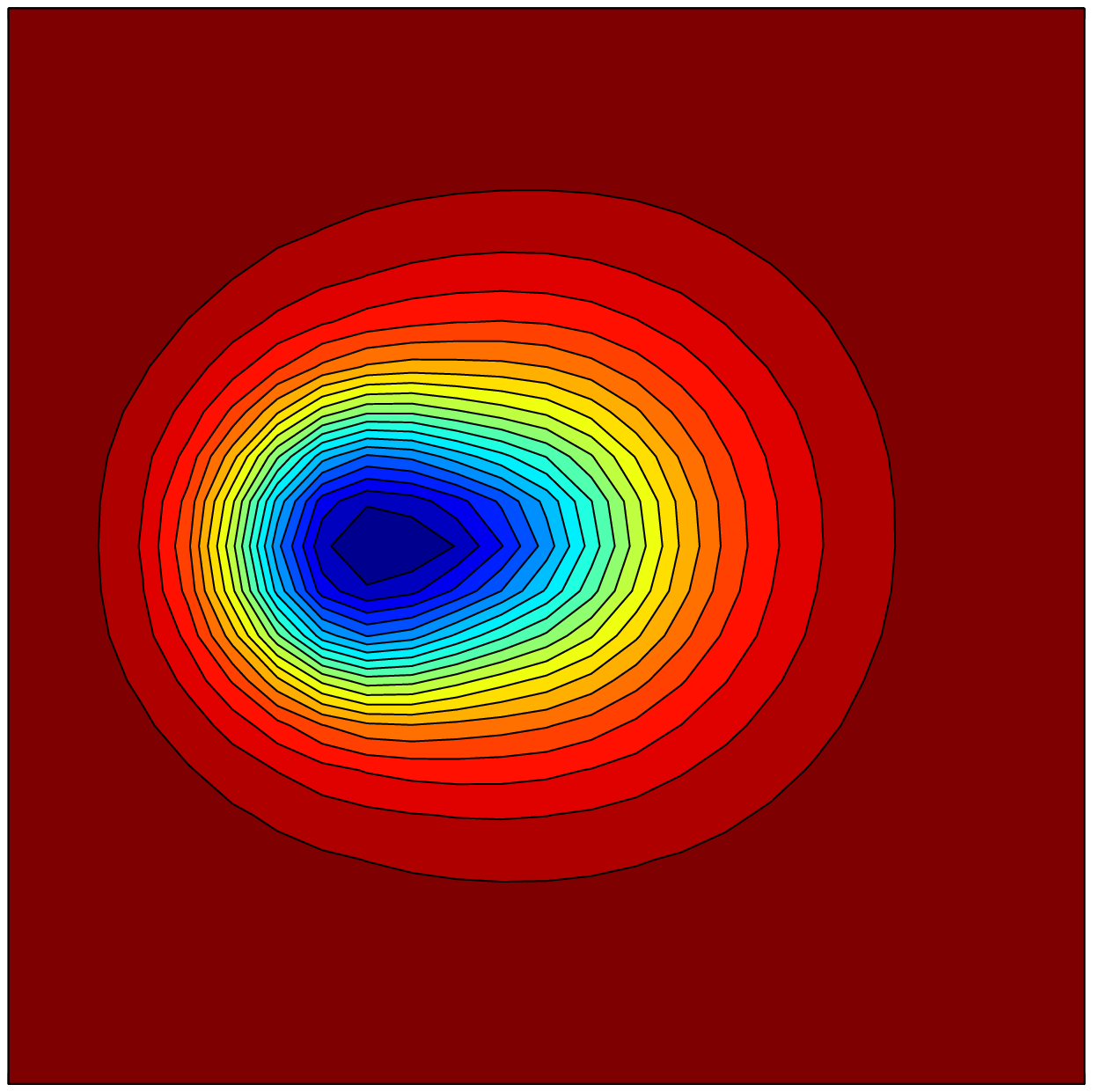}
    \end{minipage}
    }
    {
    \begin{minipage}[t]{0.31\textwidth}
    \centering
    \includegraphics[width=0.6\linewidth]{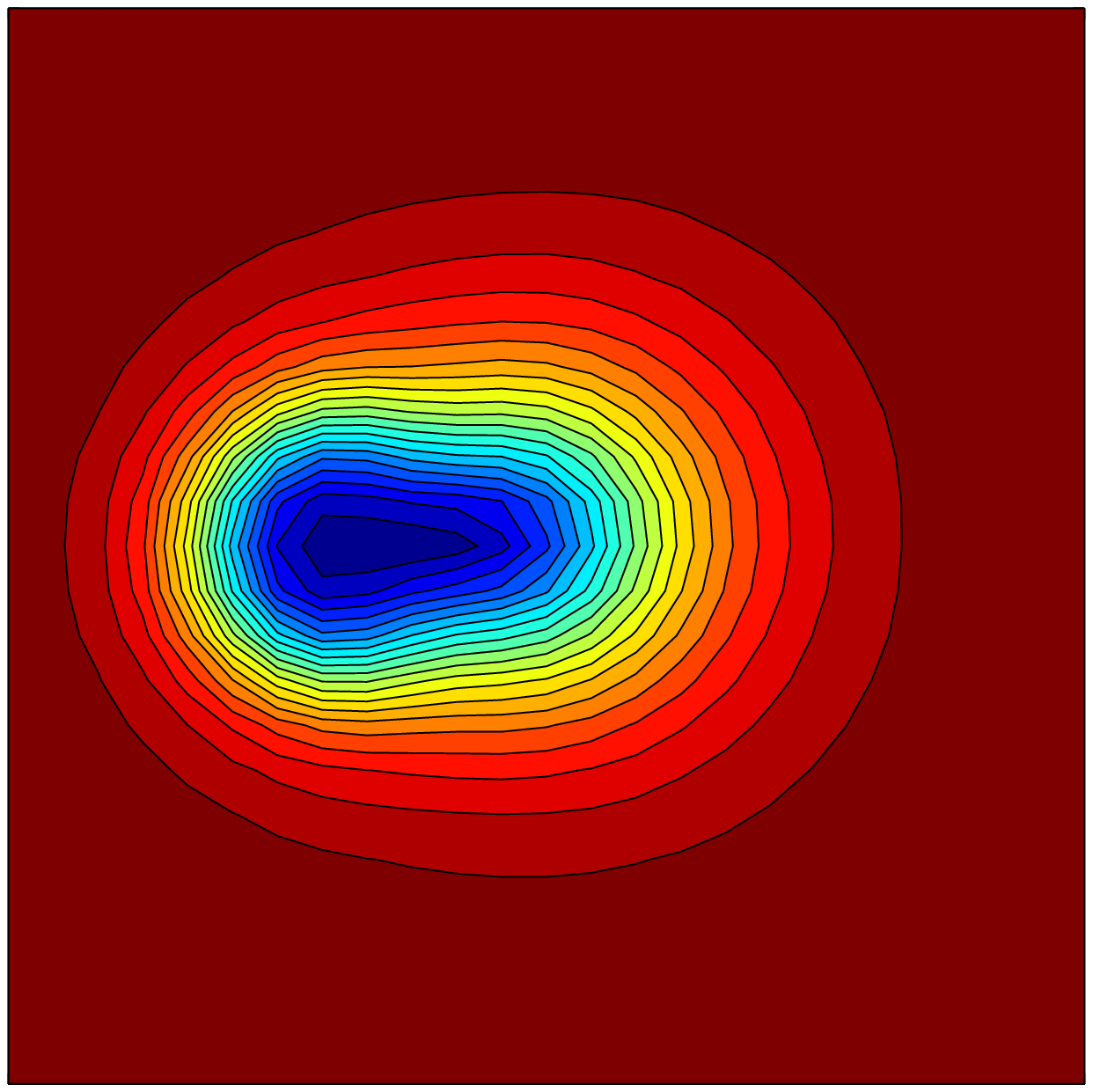}
    \end{minipage}
    }
    {
    \begin{minipage}[t]{0.31\textwidth}
    \centering
    \includegraphics[width=0.6\linewidth]{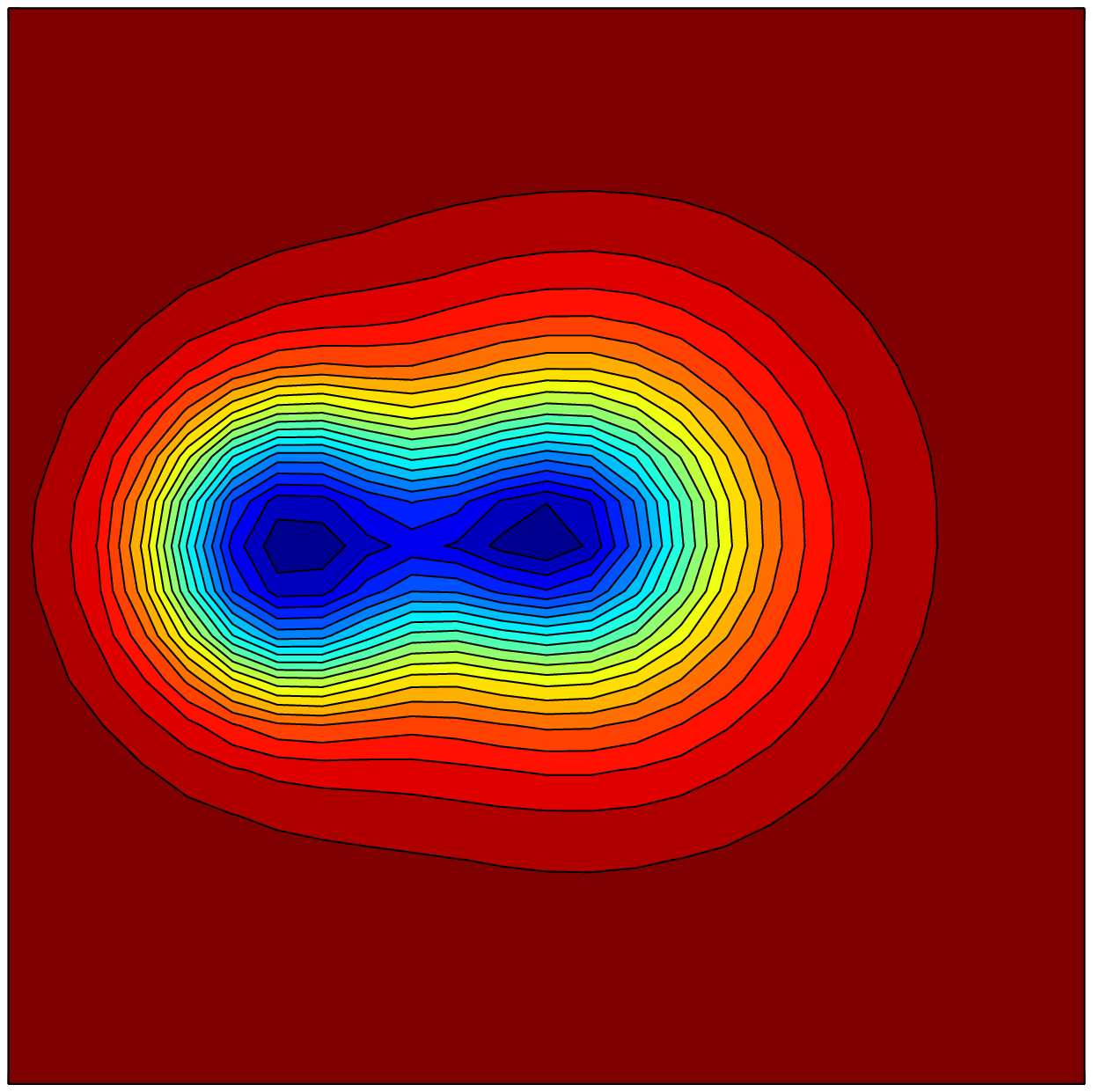}
    \end{minipage}
    }
    {
    \begin{minipage}[t]{0.31\textwidth}
    \centering
    \includegraphics[width=0.6\linewidth]{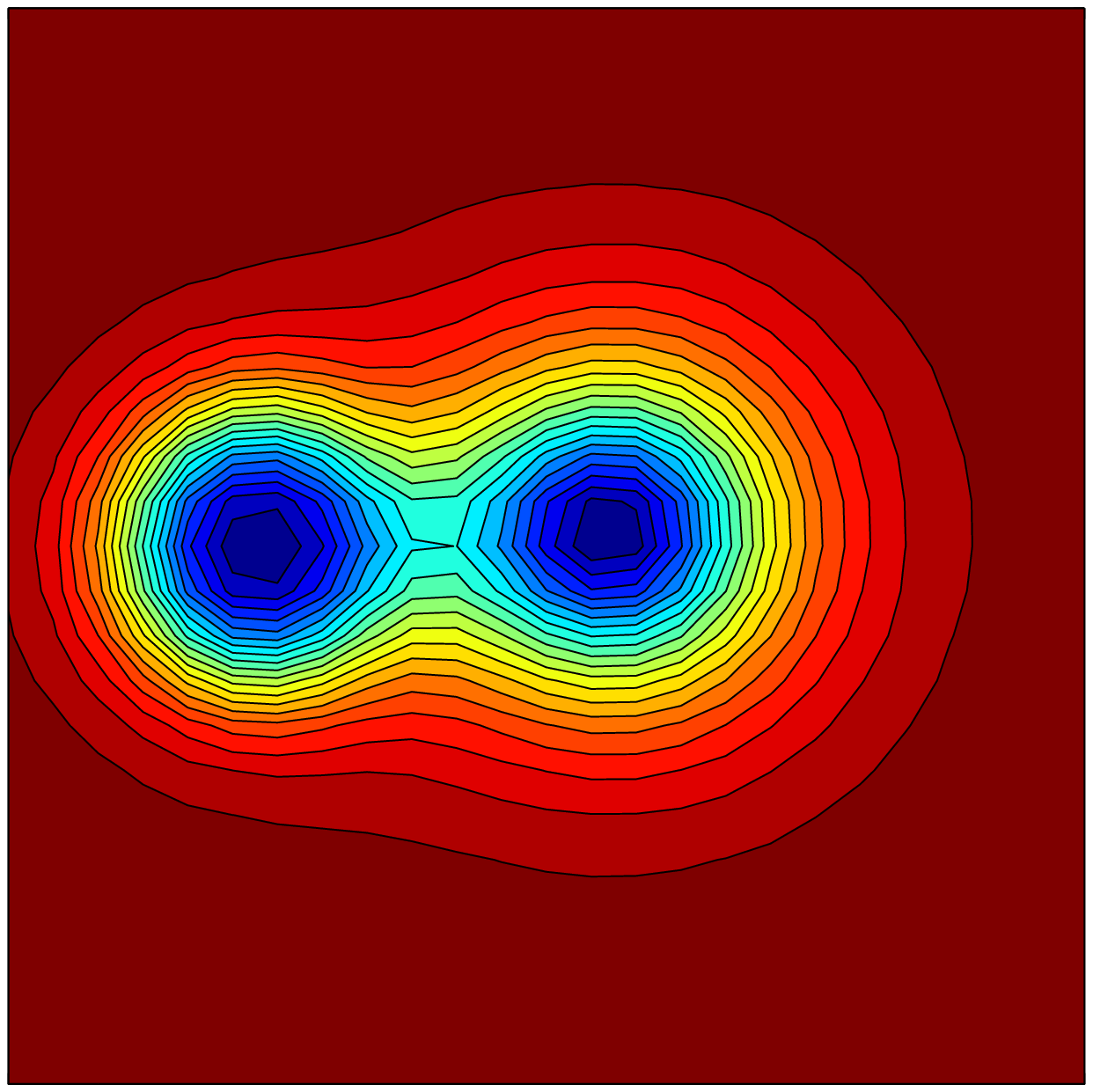}
    \end{minipage}
    }
\caption{
The same sequence of views of figure \ref{disloc} is shown
here through $z=L_z/4$ cuts of that unit cell ($L=Lz=12\xi$,$R=1.8\xi$).
For the displacement $u=3.2\xi$, the vortex line
is detached from the lower sphere.} \label{dislocutz1}
\end{figure}

\pagebreak Beyond the maximum stretch the vortex unpins from one of
the spheres, and the kinetic energy decreases for further
displacement $u$. Consequently one expects that the kinetic energy
undergoes a minimum because following the detachment of the first
pinning center from the vortex core the kinetic energy must rise
again. Notice that in case of the nucleation of two independent
surfaces from a single one the deflection of the order parameter
must be taken into account in each one of the new nucleated
surfaces. This nucleation leads to an extra growth of the kinetic
energy because there is a gradient of the order parameter in the
interfaces. The total interface area has grown because there are two
surfaces instead of just one after the nucleation. Here the two
interfaces are the $z=L_z/4$ pinning sphere and the other one formed
by the $z=3L_z/4$ pinning center together with the vortex core. The
depinning transition is shown in figure \ref{disloc} that shows a
sequence of increasing $u$ displacements that lead to the depinning
of the vortex line from the zigzag path. The same sequence is shown
again through contour lines along the two planes that cross the pinning spheres in half,
$z= L_z/4$(figure \ref{dislocutz1}) and $3L_z/4$(figure \ref{dislocutz1}).
%
%
%
%
\begin{figure}[!ht]
{
\includegraphics[width=0.6\linewidth]{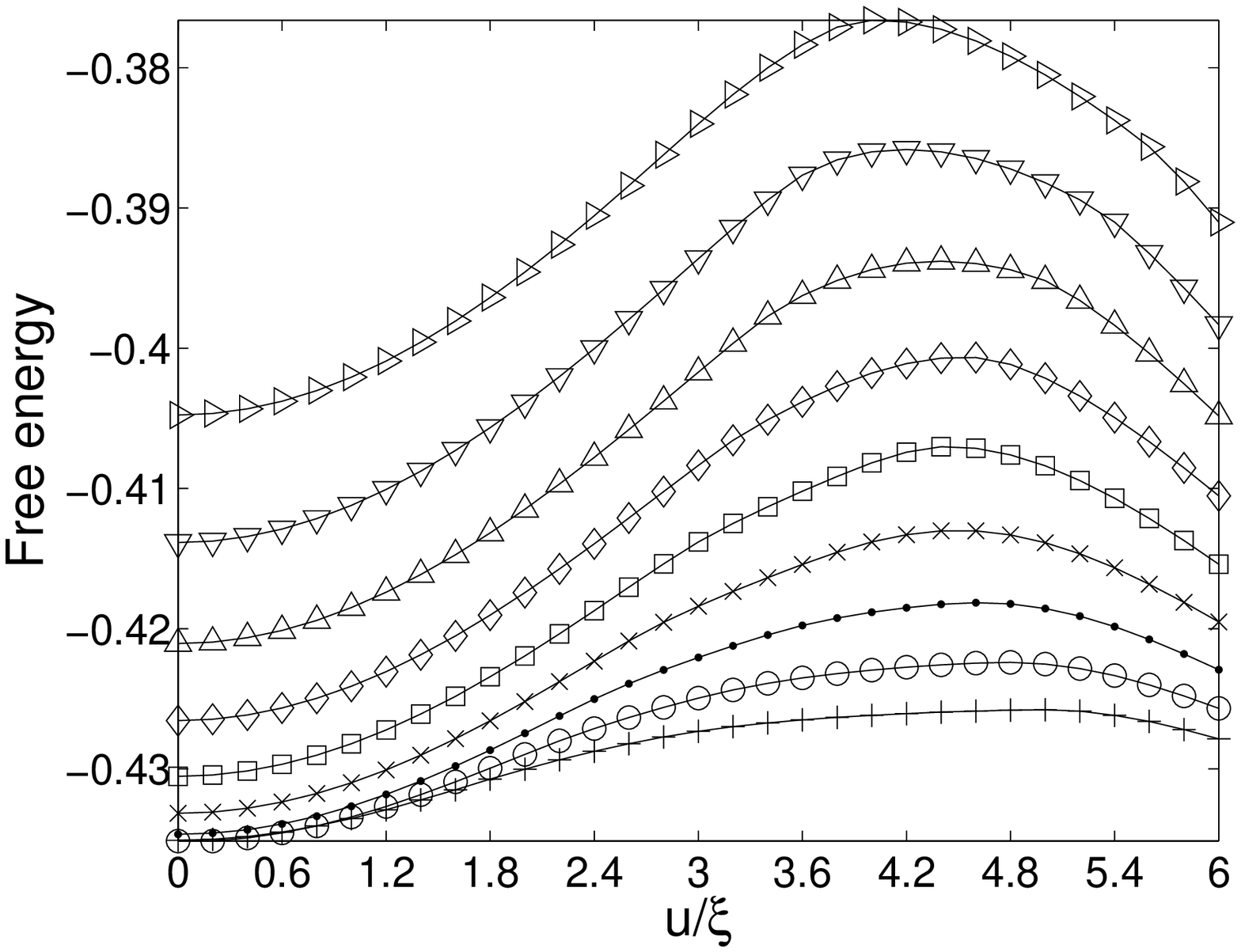}
} \caption{The free energy versus the pinning sphere dislocation for
different radii, all obtained for a cubic unit cell with
$L_z=L=12\xi$. The symbols $+$, $\circ$ , $\bullet$, $\times$,
$\sqcap$, $\diamond$, $\bigtriangleup$, $\bigtriangledown$ and
$\triangleright$ mean the radius $R/\xi$ equal to $1.2$, $1.4$,
$1.6$, $1.8$, $2.0$, $2.2$, $2.4$, $2.6$ and $2.8$ respectively. The
line of critical displacement $u_c$ is also shown here. The vortex
remains pinned to the zigzag of defects only for $u<u_c$.}
\label{fopt_u}
\end{figure}
\begin{figure}[!b]
\includegraphics[width=0.6\linewidth]{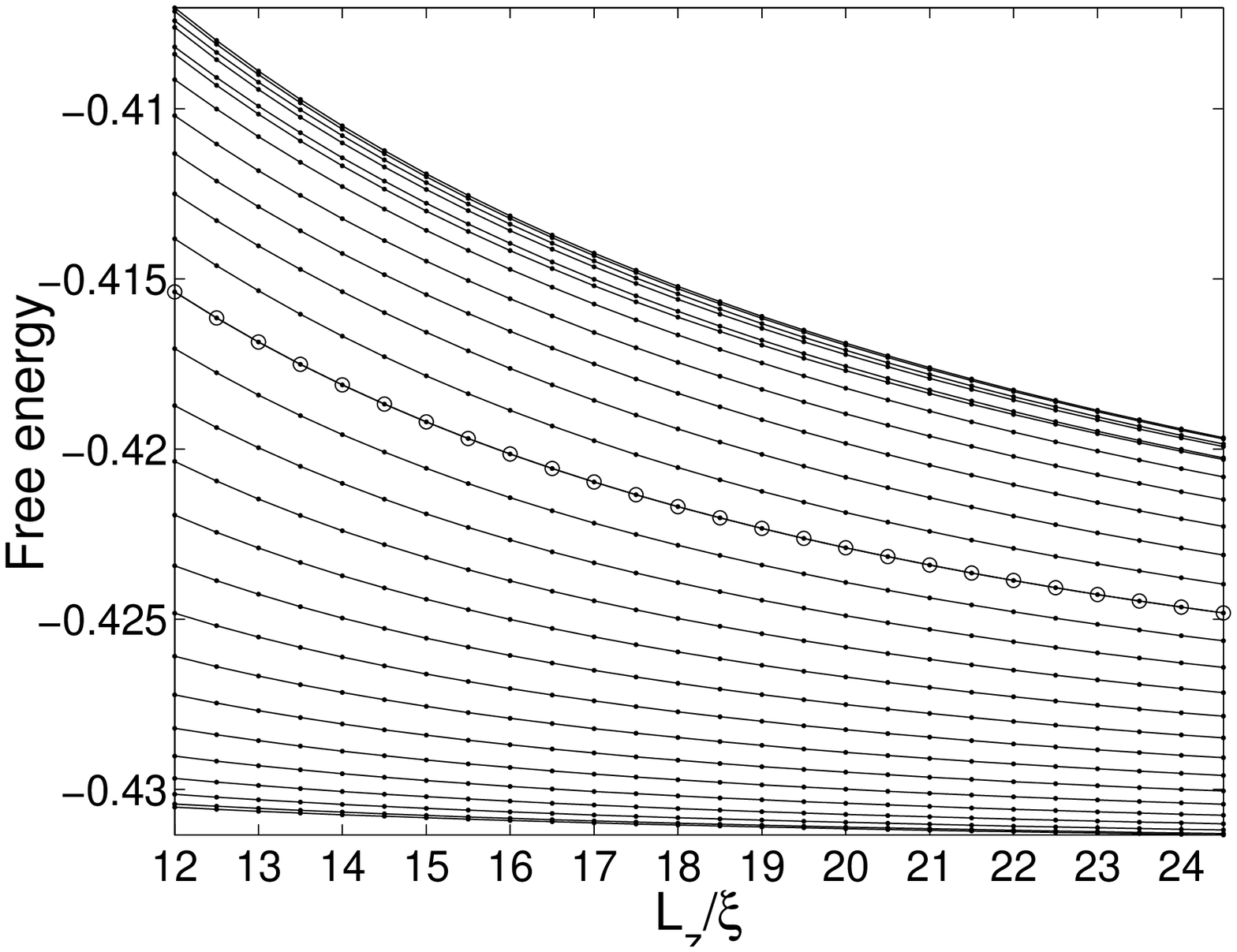}
\caption{Several $F(u,Lz,R=2.0\xi) \times Lz$ curves, associated to
different $u$ displacements, are shown here. The arrow indicate the
ascending values of $u$, ranging $0.0$ to $6.0\xi$ with a step of
$0.2\xi$.  For reference the $u=2.8\xi$ curve has its points
magnified here.} \label{fopt_lz}
\end{figure}

\begin{figure}[!t]
 \centering
 {
 \begin{minipage}[t]{0.45\textwidth}
 \includegraphics[width=\linewidth]{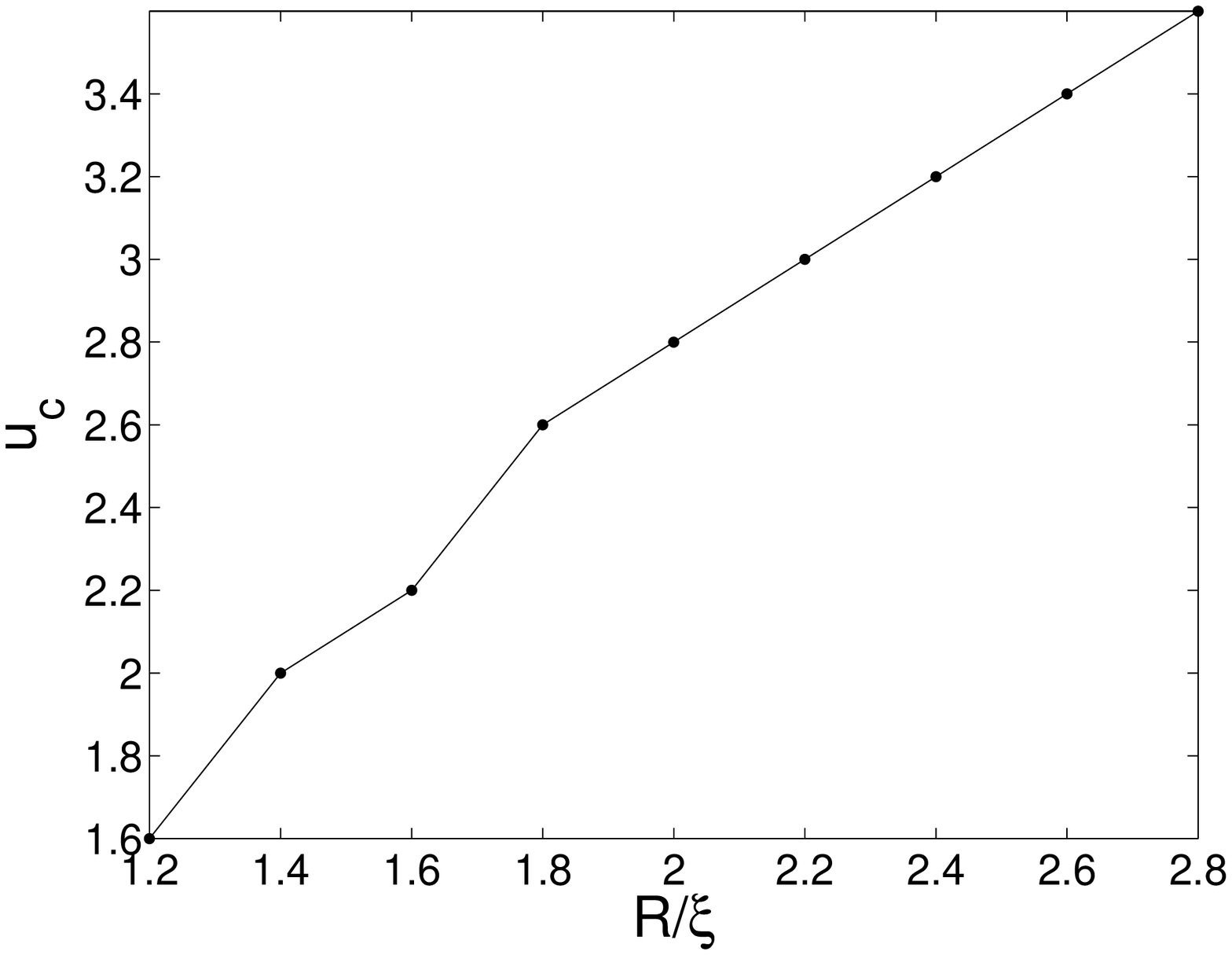}
 \end{minipage}
 }
 {
 \begin{minipage}[t]{0.45\textwidth}
 \includegraphics[width=\linewidth]{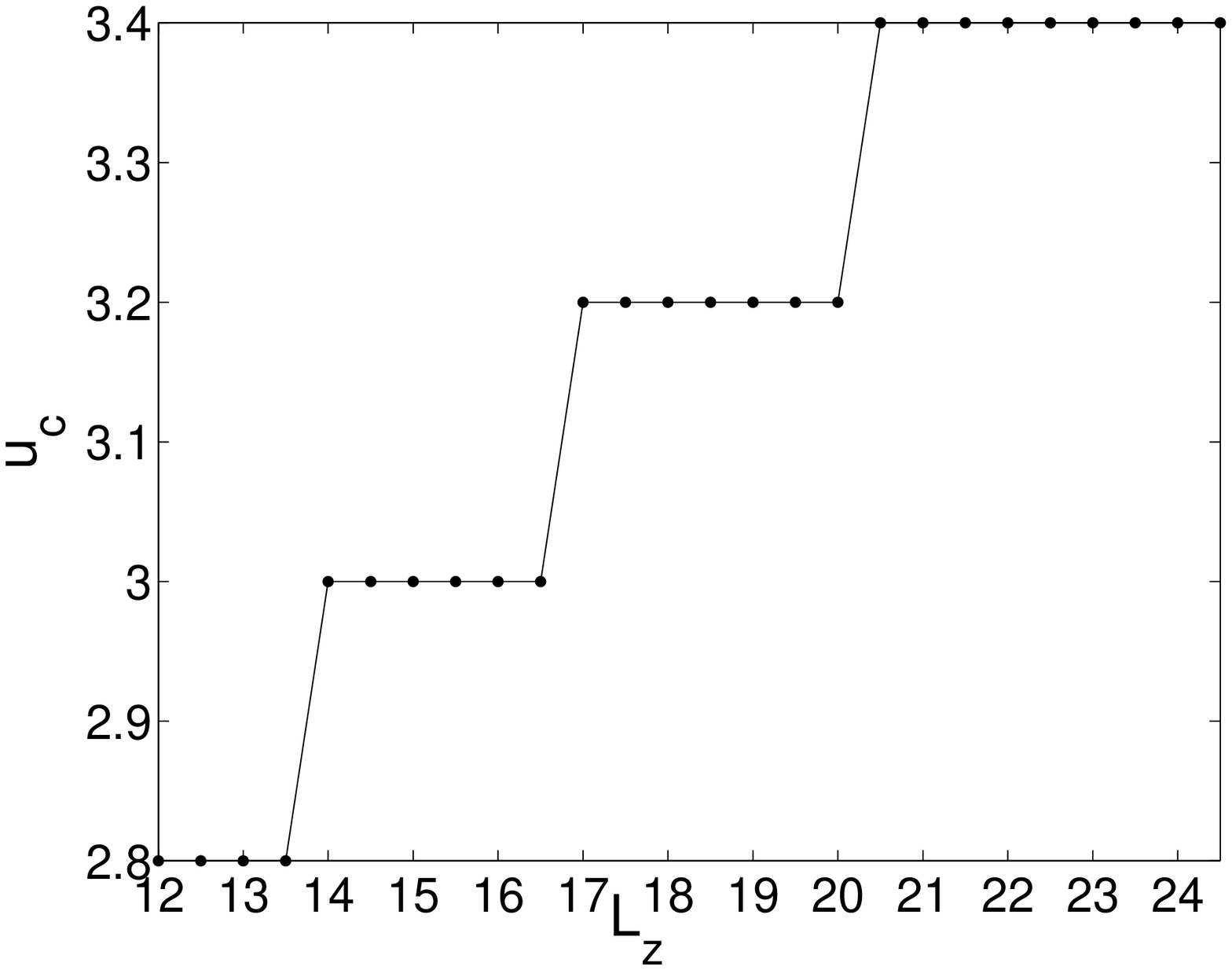}
 \end{minipage}
 }\caption{In the left side, the dependence of $u_c$ with the radius of the sphere is shown ($L=L_z=12\xi$).
In the right side, the dependence of  $u_c$ with the height $Lz$ is
presented for a defect of radius $R=2.0\xi$.} \label{depinning}
\end{figure}

%
The behavior of the free energy density versus the displacement $u$
is shown in figure \ref{fopt_u} for several radii, ranging from
$R=1.2\xi$ to $2.8\xi$, and for a cubic unit cell $L=Lz=12\xi$. The
left graphic in the figure \ref{depinning} exhibits the dependency
of the critical displacement $u_c$ with the radius of the spheres.
Due to our choice of unit cell parameter $L$,
$R$ is limited to the interval $1.0\xi<R<3.0\xi$, otherwise are smaller than the vortex core
or the pinning spheres will not fit in the unit cell.

The critical displacement is dependent on the height of the unit
cell $L_z$, as shown in figures \ref{fopt_lz} and in the right side
of the figure \ref{depinning} for the case that $R=2.0\xi$. Figure
\ref{fopt_lz} shows a set of free energy density curves versus $L_z$
associated to different $u$ displacements in steps of $0.2\xi$,
ranging from $0$ to $6.0\xi$ in the crescent sequence indicated by
the arrow. The $0$ displacement curve is the lowest in energy and
the $6.0\xi$ is the largest, in agreement with the idea that large
pinning centers increase the energy since they bring
non-superconducting regions to the unit cell. The right side of
Figure \ref{depinning} shows that unit cells with increasing $L_z$
allow for larger critical displacements. The zigzag of pinning
centers is not so demanding of the vortex line for large $L_z$, and
so, we expect that large displacements are possible in these cases,
as confirmed by figures \ref{fopt_lz} and \ref{depinning}.

\section{Conclusions}
\label{Conclusions} The study of the interaction among pinning
centers and vortices is crucial to the understanding of type II
superconductors. For this reason artificially made pinning centers
are useful and many kinds have been
fabricated\cite{BP95,LGTBHP96,MBMRBTBJ98,YLBJW04,PM04} though not of
the kind considered here. In this paper the Ginzburg-Landau theory
was numerically solved ona three-dimensional mesh to study properties of
a vortex line near a zigzag of pinning centers.
The pinning centers are insulating spheres of coherence length size and the system was described by an
orthorhombic unit cell containing two displaced defects. We find
that there is a critical displacement above which the vortex line is
detached from a pinning center. Below this transition the vortex
line is pinned by both defects and above by just one. The study of
coherence length size defects may also be found useful in some other
situations. Recently it has been shown that the nucleation of such
defects is a mechanism to lower the energy of the superconducting
state\cite{DR04}.
\section{Acknowledgments}
\label{Acknowledgments}

Research supported in part by Instituto do Mil\^enio de
Nano-Ci\^encias, CNPq, and FAPERJ (Brazil).

\end{document}